\documentclass[]{article}
\usepackage{graphicx}
\usepackage{xcolor}
\usepackage{amsfonts}
\usepackage{amssymb}
\def\ligne#1{\hbox to\hsize{#1}}
\def\leurre{\noindent\leftskip0pt\small\baselineskip 10pt}
\newtheorem{thm}{\textrm{\sc Theorem}}
\newtheorem{cor}{\textrm{\sc Corollary}}
\newtheorem{lemm}{\textrm{\sc Lemma}}
\newtheorem{fig}{\textrm{Figure}}
\newtheorem{tab}{\textrm{Table}}

\newcounter{laform}
\author{Maurice {\sc Margenstern}}
\title{About Fibonacci trees. 
\begin{center} $-$ I $-$ \end{center}}
\def\Rr#1{{\color{red}{#1}}}

\def\Oo#1{{\color{orange}{#1}}}
\def\Bb#1{{\color{cyan}{#1}}}
\begin{document}
\maketitle
\begin{abstract}
In this first paper, we look at the following question: are the properties of the Fibonacci 
tree still true if we consider a finitely generated tree by the same rules but
rooted at a black node?
The direct answer is no, but new properties arise, a bit more complex than in the case
of a tree rooted at a white node, but still of interest.
\end{abstract}

\section{Introduction}
\def\zz#1{{\footnotesize\tt #1}}

   This paper investigates a question the author raised to himself a long time ago 
but he had no time to look at it. When he established the properties of the finitely
generated tree he called the Fibonacci tree, he wandered whether the properties still
hold if, keeping the generating rules, we apply them to a tree rooted at a black node.

   In section~\ref{trees}, we remind the reader the definitions about the Fibonacci
tree and its connection with two tilings of the hyperbolic plane we call {\bf pentagrid}
and {\bf heptagrid}. In section~\ref{fibo}, we consider the properties which rely on the
representations of the positive integers as sums of distinct Fibonacci numbers and we look
at the question raised in the abstract. We show that the properties attached to the 
Fibonacci tree rooted at a white node are no more true for the similar tree rooted at
a black node. However, other properties can be established, more complicate in that
setting. We also show the connection of the Fibonacci tree rooted at a black node
with the pentagrid and the heptagrid. In section~\ref{golden} we consider another
increasing sequence of positive numbers which we call the golden sequence which
can also be attached to the Fibonacci trees, both the tree rooted at a white node and
the other rooted at a black one. Again, the nice properties which connect the golden
sequence with the Fibonacci tree rooted at a white node are no more true when it
is rooted at a black node. However, other properties occur, more complicate than the
previous ones, but still worth of interest.
   
   Section~\ref{conclude} concludes the paper with open questions regarding the 
generalization of these results. It might be the goal of other papers.

\section{The Fibonacci trees}\label{trees}

   In sub section~\ref{fibodef}, we remind the reader the definition of the Fibonacci
tree as well as different variations about it which were investigated by the author
in~\cite{mmJUCStools}. In sub section~\ref{preferred}, we remind him/her the
numbering of the nodes and the important properties of their Fibonacci representations,
in particular the connections of the representation of the sons of a node with the
representation of that node. In sub section~\ref{tilings}, we remind the connection of the
Fibonacci tree with the pentagrid and with the heptagrid.

\subsection{Fibonacci trees rooted at a white node}\label{fibodef}

    We call {\bf Fibonacci tree} the finitely generated tree with two kinds of nodes,
{\bf black} nodes and {\bf white} ones whose generating rules are:
\vskip 5pt
\ligne{\hfill$B\rightarrow BW$ and $W\rightarrow BWW$.\hfill (1)\hskip 10pt}
\vskip 5pt

When the root of the tree is a white, black node, we call such a Fibonacci tree a 
{\bf white}, {\bf black} respectively, {\bf Fibonacci tree}. In this whole section,
we consider white Fibonacci trees only. Black Fibonacci trees will be studied in
Sections~\ref{fibo} and~\ref{golden}. The {\bf status} of a node says whether it is black
or white. 
 
   The connection with Fibonacci numbers first appear when we count the number of nodes
which lay at the same level of the tree. For a white Fibonacci tree, we have the
following property:

\begin{thm}{\rm\cite{mmJUCStools}}\label{tfibolevelw}
In a white Fibonacci tree, the level of the root being~$0$, the number of nodes 
on the level~$k$ of the tree is $f_{2k+1}$ where $\{f_n\}_{n\in\mathbb N}$ is the 
Fibonacci sequence with initial conditions \hbox{$f_0=f_1=1$}.
\end{thm}

   As in~\cite{mmJUCStools}, let us number the nodes starting from~1 given to the root
and then, from level to level and, on each level, from left to right. From now on, we 
identify a node with the number it receives in the just described way.
It is known
that any positive integer~$n$ can be written as a sum of distinct Fibonacci numbers,
the terms of the Fibonacci sequence considered in Theorem~\ref{tfibolevelw}:
\vskip 5pt
\ligne{\hfill$n=\displaystyle{\sum\limits_{i=1}^k a_if_i}$ with $a_i\in\{0,1\}$.\hfill
(2)\hskip 10pt}

\subsection{The preferred son property}\label{preferred}

The $a_i$ digits which occur in~(1) are not necessarily unique for a given~$n$. 
They can be made unique
by adding the following condition~: for \hbox{$0<i<k$}, if \hbox{$a_i=1$}, then
$a_{i-1}=0$. If we write the $a_i$'s of~(1) as a word \hbox{\bf a$_1$..a$_k$},
the condition for uniqueness of the representation says that in that word, the pattern
{\bf 11} never occurs. We call {\bf code} of~$n$ the unique word attached to~$n$ by~(1) 
when the pattern {\bf 11} is ruled out. We also say code of~$\nu$ for the node of the white
Fibonacci tree whose number is~$\nu$, and we write $[\nu]$ for the code of~$\nu$.
Formula~(2) allows us to restore $\nu$ from its code $[\nu]$. We shall write
\hbox{$\nu = ([\nu])$}. From Theorem~\ref{tfibolevelw}, we get:

\begin{cor}{\rm\cite{mmJUCStools}}\label{cheadlevelw}
In a white Fibonacci tree, the leftmost node of the level~$k$ has the number
$f_{2k}$ and has the code {\bf 10$^{2k-1}$} for positive~$k$. The rightmost node on the same
level has the number $f_{2k}$$-$$1$ and has the code {\bf 1$($01$)^k$}.
\end{cor}

   We can state the following property:

\begin{thm}{\rm\cite{mmJUCStools}}\label{tpreferred} 
In a white Fibonacci tree, for any node~$\nu$ we have that among its sons
a single one has $[\nu]00$ as its code. That son is called the {\bf preferred son}
of~$\nu$. If the node is black, its preferred son is its black son, if the node
is white, its preferred son is its white son in between its black son and the other
white one.
\end{thm}

Figure~\ref{ffiboblanc} illustrates the properties stated in the theorem. The reader
may easily check them.

\vskip 10pt
\vtop{
\ligne{\hfill
\includegraphics[scale=0.35]{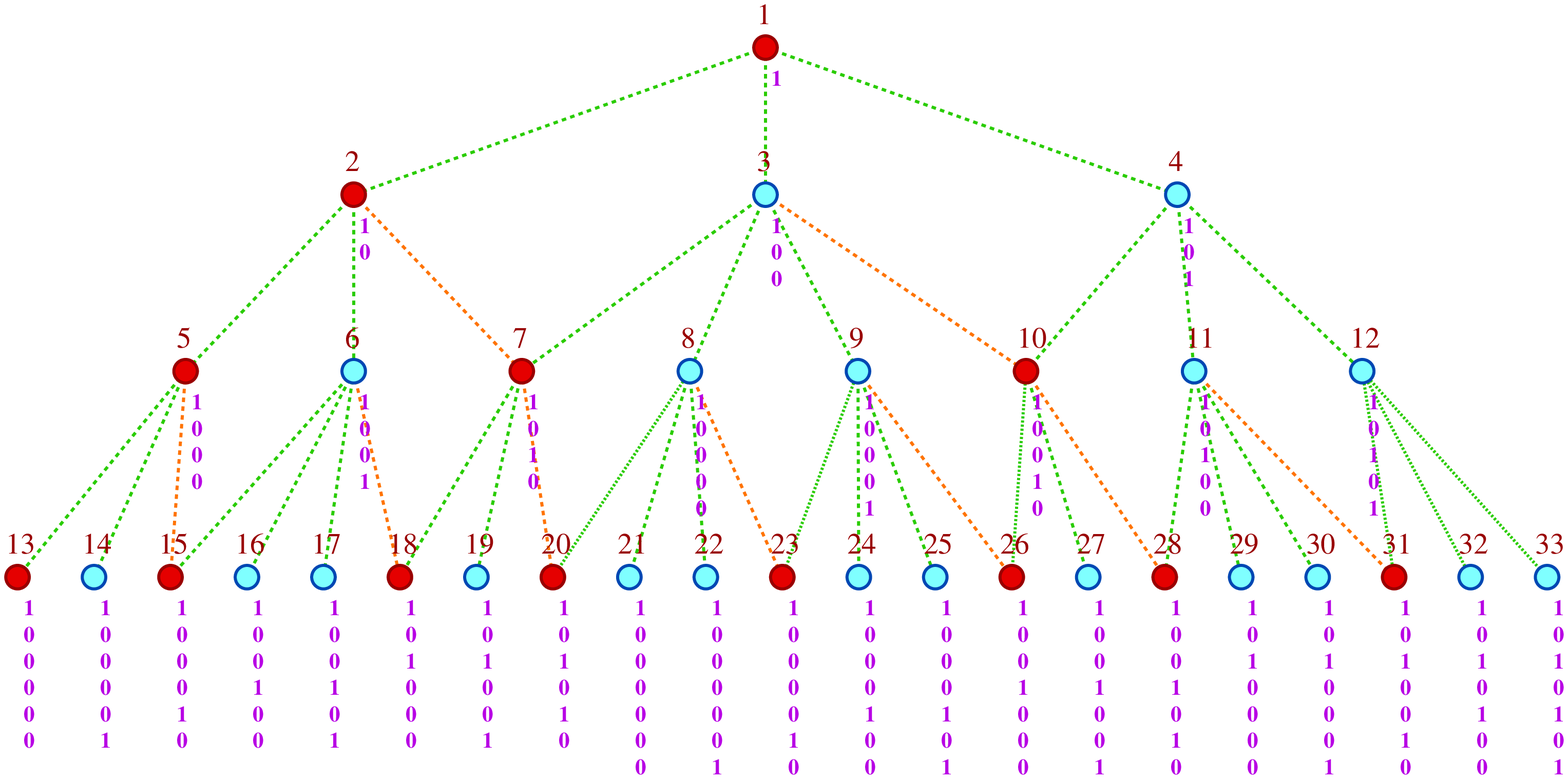}
\hfill}
\vspace{-10pt}
\ligne{\hfill
\vtop{\leftskip 0pt\parindent 0pt\hsize=300pt
\begin{fig}\label{ffiboblanc}
\leurre
The white Fibonacci tree. Representation of the first three levels of the tree.
In blue, the white nodes, in red, the black ones. Above a node, its number, under it,
its code. Note that the preferred son property is true as stated in 
Theorem~{\rm\ref{tpreferred}}. The green edges connect a node to its sons, while the
orange ones connect a node~$n$ to the leftmost son of the node~$n$$+$$1$ when that latter
node is present on the considered level.
\end{fig}
}
\hfill}
}
\vskip 5pt
\subsection{Connection of the Fibonacci tree with the pentagrid and with the
heptagrid}~\label{tilings}

   As mentioned in the introduction, the white Fibonacci tree is connected with the
pentagrid and the heptagrid, two tilings of the hyperbolic plane. The pentagrid is the
tessellation $\{5,4\}$, which means that the tiling is generated by the reflection of
a basic polygon in its sides and the recursive reflections of the images in their sides,
where the basic polygon is the regular convex pentagon with right angles. That polygon
lives in the hyperbolic plane, not in the Euclidean one. Similarly, the heptagrid is the 
tessellation $\{7,3\}$ which is generated in a similar way where the basic polygon is the 
regular convex heptagon with $\displaystyle{{2\pi}\over3}$ as its vertex angle. Again, 
such a polygon lives in the hyperbolic plane and not in the Euclidean plane. 
Figure~\ref{fpentahepta} illustrates the pentagrid, left hand side, and the heptagrid,
right hand side. 

\vskip 10pt
\vtop{
\ligne{\hfill
\includegraphics[scale=0.6]{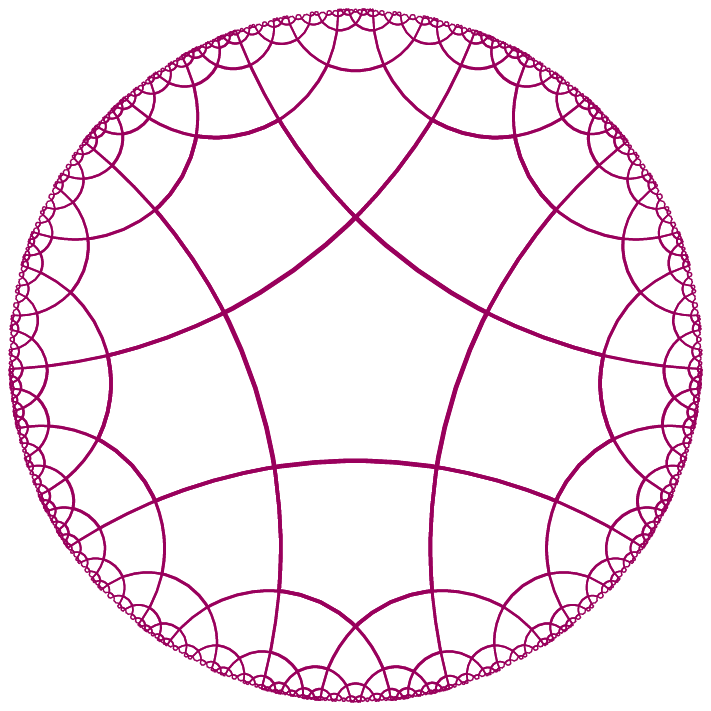}
\includegraphics[scale=0.6]{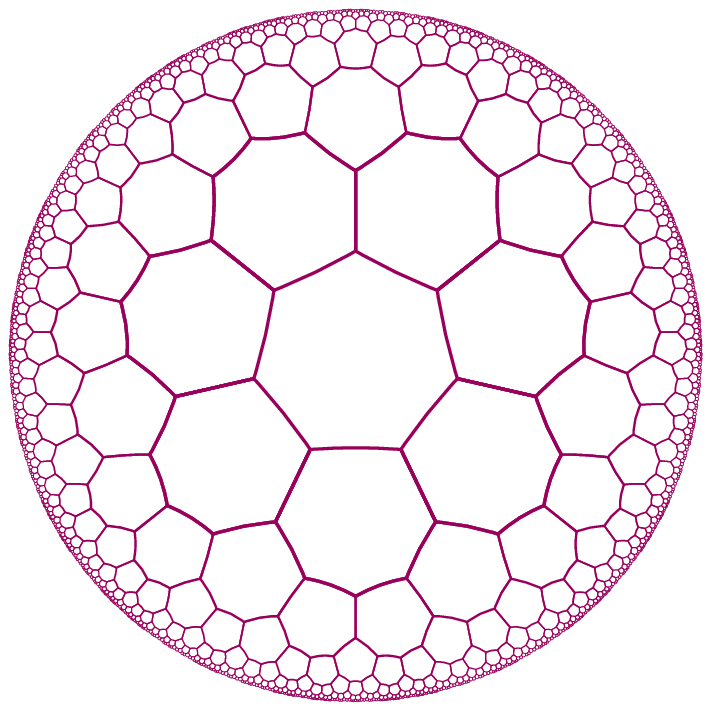}
\hfill}
\vspace{-10pt}
\ligne{\hfill
\vtop{\leftskip 0pt\parindent 0pt\hsize=300pt
\begin{fig}\label{fpentahepta}
\leurre
The tilings generated by the white Fibonacci tree.
To left, the pentagrid, to right, the heptagrid.
\end{fig}
}
\hfill}
}
\vskip 5pt
Figure~\ref{fspanpentahepta} illustrates how the white Fibonacci tree generates
the considered tilings. In both tilings, the tiles of a {\bf sector}, can be put in
bijection with the nodes of the tree. In the case of the pentagrid, such a sector is
a quarter of the plane: it is delimited by two perpendicular half-lines stemming from
the same vertex~$V$ of a tile~$\tau$ and passing through the other ends of the edges
of~$\tau$ sharing~$V$.
\vskip 10pt
\vtop{
\ligne{\hfill
\includegraphics[scale=0.45]{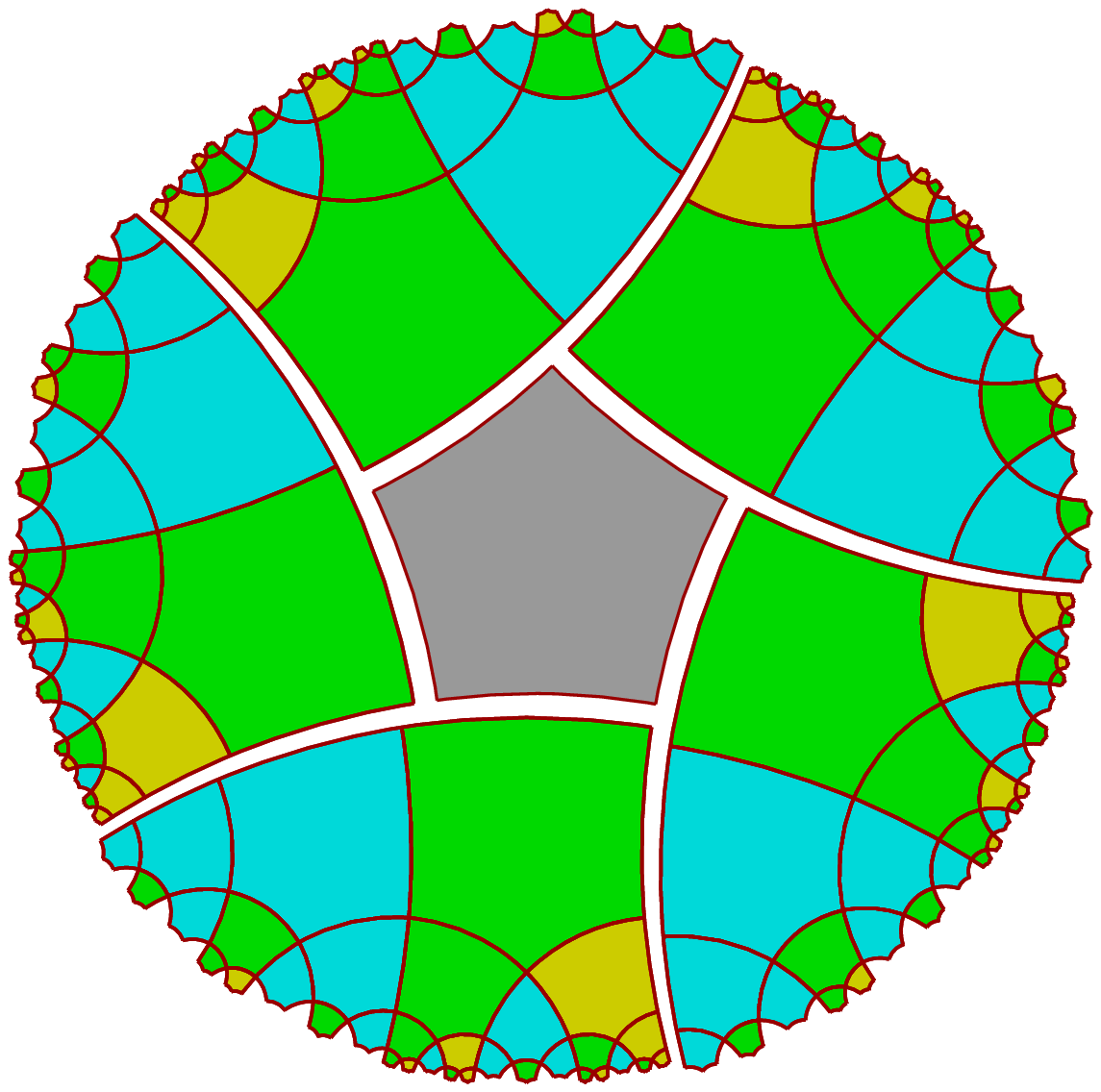}
\includegraphics[scale=0.445]{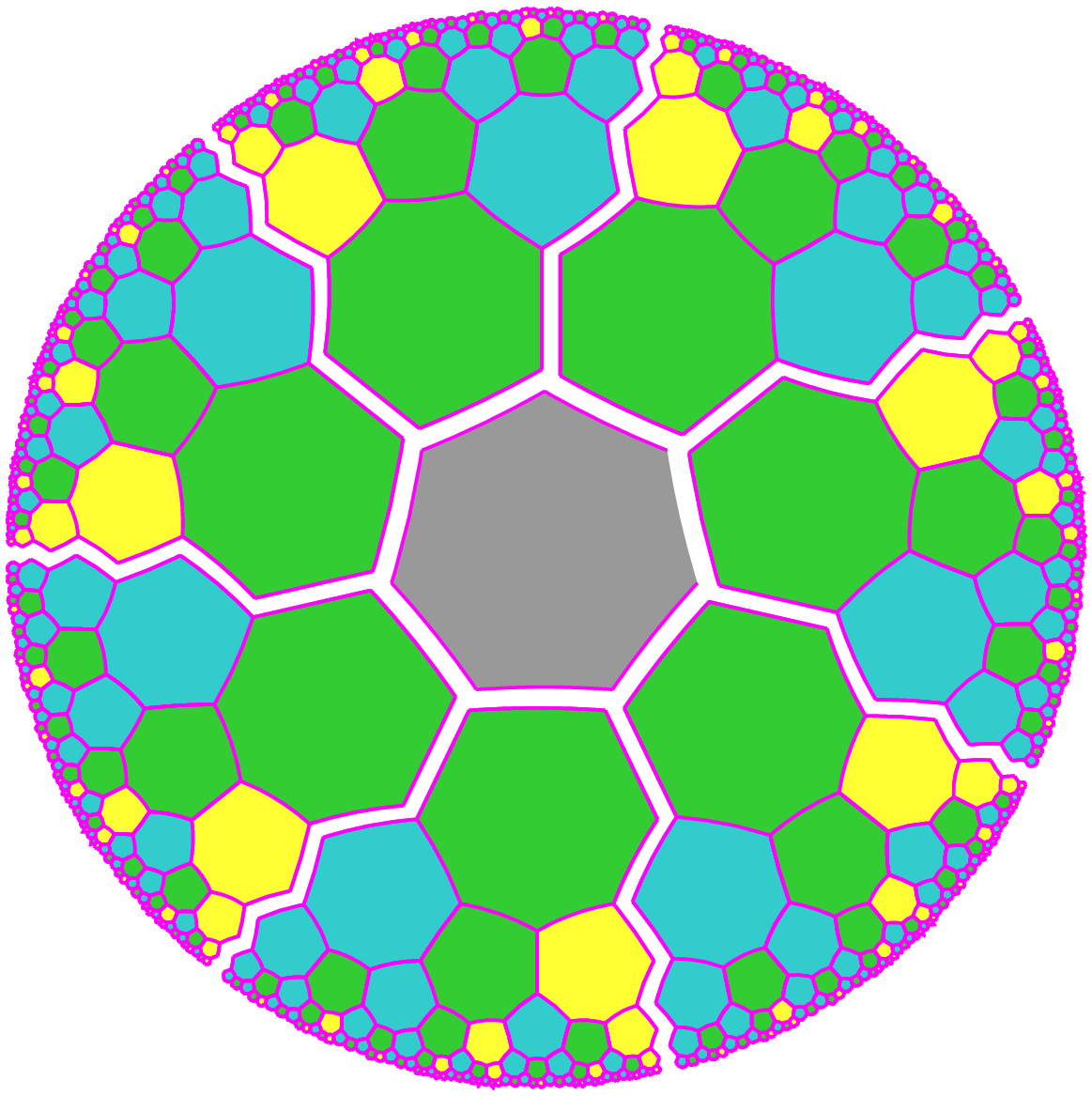}
\hfill}
\vspace{-10pt}
\ligne{\hfill
\vtop{\leftskip 0pt\parindent 0pt\hsize=300pt
\begin{fig}\label{fspanpentahepta}
\leurre
How the white Fibonacci trees generate the pentagrid and the heptagrid:
each isolated sectors in the above figures is spanned by the white Fibonacci tree.
The colours of the tile show the Fibonacci structure: blue tiles are the black nodes,
green and yellow tiles are the white ones. Yellow tiles are the second white son of a
white node.
\end{fig}
}
\hfill}
}
\vskip 5pt
The definition of the sector is more delicate in the case of the heptagrid. The sector
is also defined by half-lines which are, this time, issued from the mid-point of an 
edge~$\eta$ and those half-lines pass through the mid-points of two consecutive sides of 
a tile sharing~$V$ as a vertex, $V$ being also an end of~$\eta$. The reader is
referred to~\cite{mmbook1} for proofs of the just mentioned properties. Accordingly,
as shown on the figure, five sectors allow us to locate tiles in the pentagrid and
seven sectors allows us to perform the same thing in the heptagrid. From now on,
we call {\bf tile $\nu$} the tile attached to the node~$\nu$ of such a Fibonacci tree
we assume to be fixed once and for all. We also say that $[\nu]$ is the code of
the tile~$\nu$.

The preferred son property allows us to compute in linear time with respect to the code
of a node~$\nu$ the codes of the nodes attached to the tiles which share a side 
with the tile~$\nu$. Such tiles are called the {\bf neighbours} of~$\nu$. 
Theorem~\ref{tpreferred} also allows us to compute in linear time with respect to
$[\nu]$ a shortest path in the tiling, leading from the tile~$\nu$ to tile~1.

\section{When Fibonacci numbers are used}\label{fibo}

   So far, we mentioned the properties of the white Fibonacci tree. Let us look at the
following problem with the black Fibonacci tree. In Sub section~\ref{bfibo}, we look at the
analog of Theorem~\ref{tfibolevelw} and its Corollary~\ref{cheadlevelw}. In 
Sub section~\ref{bsucc}, we investigate the analog of Theorem~\ref{tpreferred}. In
Sub section~\ref{bpentahepta}, we look at the connection of the black Fibonacci tree
with the pentagrid and with the heptagrid.

\subsection{The black Fibonacci tree and Fibonacci numbers}\label{bfibo}

   We can prove an analog of Theorem~\ref{tfibolevelw}:

\begin{thm}\label{tfibolevelb}
In a black Fibonacci tree, the level of the root being~$0$, the number of nodes 
on the level~$k$ of the tree is $f_{2k}$ where $\{f_n\}_{n\in\mathbb N}$ is the 
Fibonacci sequence with initial conditions \hbox{$f_0=f_1=1$}.
\end{thm}

\begin{cor}{\rm\cite{mmJUCStools}}\label{ctaillevelb}
In a black Fibonacci tree, the rightmost node of the level~$k$ has the number
$f_{2k+1}$ and has the code {\bf 10$^{2k}$} for non negative~$k$.
\end{cor}

   The property stated in Theorem~\ref{tfibolevelb} was noted in a paper by Kenichi Morita, 
but we shall see in 
Section~\ref{golden} that both Theorems~\ref{tfibolevelw} and~\ref{tfibolevelb} 
simply come from the fact that the same generating rules are applied
to the trees and that the initial levels contain 1 and 3 nodes for the white tree
while they contain 1 and 2 nodes for the black one.
The latter
properties are in some sense symmetric: in the white Fibonacci tree the levels
are Fibonacci numbers with odd index so that the cumulative sum up to the current level
plus one is a Fibonacci number with even index and in the black Fibonacci tree the situation
is opposite: the levels give rise to Fibonacci numbers with even index whose cumulative
sum is precisely a Fibonacci number with odd index. It is the reason why in the white
tree, the Fibonacci numbers with even index occur at the head of a level why in the black
one the Fibonacci numbers with odd index occur at the tail of a level.

\subsection{The black Fibonacci tree and a successor property}\label{bsucc}

   The next question is: is Theorem~\ref{tpreferred} true for the black node if we number
its nodes in the same way as in the white tree and if we attach to the nodes the codes
of their numbers? The answer is no if we stick to the same statement and it is yes if
we change the question by asking whether there is a rule to define the connection
of a node~$\nu$ with the one whose code is~$[\nu]00$. Such a situation can be
suspected by comparing Theorem~\ref{tfibolevelw} with Theorem~\ref{tfibolevelb}
as well as Corollary~\ref{cheadlevelw} with Corollary~\ref{ctaillevelb}. 

\vskip 10pt
\vtop{
\ligne{\hfill
\includegraphics[scale=0.35]{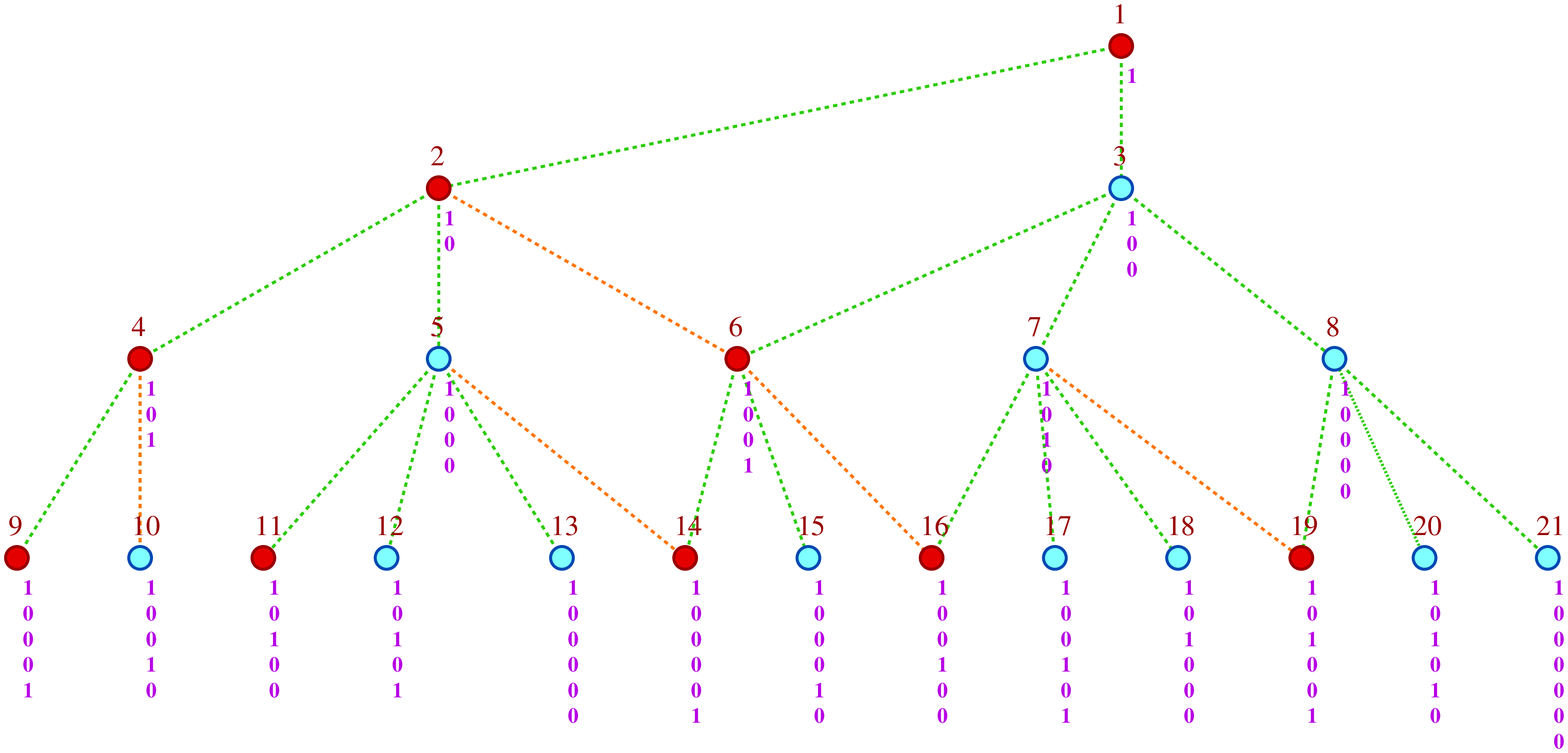}
\hfill}
\vspace{-10pt}
\ligne{\hfill
\vtop{\leftskip 0pt\parindent 0pt\hsize=300pt
\begin{fig}\label{ffibonoir}
\leurre
The black Fibonacci tree. The same convention about colours of the nodes and of the edges
between nodes as in Figure~{\rm\ref{ffiboblanc}} is used. 
We can see that the preferred son property as
stated in Theorem~{\rm\ref{tpreferred}} is not true in the present setting.
\end{fig}
}
\hfill}
}
\vskip 5pt
Figure~\ref{ffibonoir} shows us a first fact: the preferred son property is not observed
in the black Fibonacci tree. We can see that for any black node~$\nu$, none of its sons
has the code {\bf [$\nu$]00}. More other, for the white nodes~$\nu$ whose code can be 
written {\bf [$\mu$]01}, the code {\bf [$\mu$]0100} is that of the leftmost son of the 
node~$\nu$+1.

\def\bbzz{{\bf b00}}
\def\bbzu{{\bf b01}}
\def\wwzz{{\bf w00}}
\def\wwzu{{\bf w01}}
\def\wwuz{{\bf w10}}
   We can now state the property which holds in the black Fibonacci tree. As the preferred
son property is no more observed with the exception of a few white nodes,
we say that the {\bf successor} of the node~$\nu$ is the node whose code is
{\bf [$\nu$]00}. Theorem~\ref{tpreferred} says that in a white node, the successor of a
node occurs among its sons with a precise rule depending on the colour of the node.
In order to formulate the property, we need to define the {\bf end} of the code of a node.
The last two digits of the code of a node is either {\bf 00}, {\bf 01} or {\bf 10}. In
that latter case, we know that in fact we can write the last three digits as {\bf 010}
as the pattern {\bf 11} is ruled out. With two kinds of nodes, this would give us
six classes of nodes. In fact we have five of them only, which can be written
\bbzz, \bbzu, \wwzz, \wwzu{} and \wwuz, which we call the {\bf types}
of the nodes. The type of a node mixes its status with the ending of its code. 
Accordingly, we can see that the type \bbzu{} defines an empty class, a property we 
shall check. For a node~$\nu$, $succ(\nu)$ is the number of its successor and
$s_r(\nu)$, $s_\ell(\nu)$ is the number of its rightmost, leftmost son respectively.
Note that we always have:
\vskip5pt
\ligne{\hfill $s_r(\nu)$+1 = $s_\ell(\nu$+$1)$.\hfill (3)\hskip 10pt}

\begin{thm}\label{tsuccb}
In the black Fibonacci tree, we have \hbox{$succ(\nu)=s_r(\nu)$$+$$1$} if $\nu$ 
is a black node or a white node of type \wwzu. For the types of white node
\wwzz{} and \wwuz{}, we have \hbox{$succ(\nu) = s_r(\nu)$}. We have six 
rules giving the types of the sons of a node according to its type:
\vskip 5pt
\ligne{\hfill
$\vcenter{\vtop{
\leftskip 0pt\parindent 0pt\hsize=150pt
\ligne{\bf \bbzz{} $\,\,\,\,\,\rightarrow$ \bbzu{} - \wwuz\hfill}
\ligne{\bf \bbzu{} $\,\,\,\,\,\rightarrow$ \bbzu{} - \wwuz\hfill}
\ligne{\bf \wwzz{} $\,\,\,\rightarrow$ \bbzz{} - \wwzu{} - \wwzz\hfill}
\ligne{\bf \wwzz$\ast$ $\rightarrow$ \bbzu{} - \wwuz{} - \wwzz$\ast$\hfill}
\ligne{\bf \wwzu{} $\,\,\,\rightarrow$ \bbzz{} - \wwzu{} - \wwuz\hfill}
\ligne{\bf \wwuz{} $\,\,\,\rightarrow$ \bbzz{} - \wwzu{} - \wwzz\hfill}
}}$
\hfill}
\vskip 5pt
\noindent
where the type \wwzz$\ast$ indicates a node of the form $f_{2k+1}$.
\end{thm}
  
\noindent
Proof. We prove the theorem by induction on all the properties listed in it. We note that 
the properties are trivially observed on Figure~\ref{ffibonoir} for the root and its sons.
Accordingly, we assume the property to be true for all nodes \hbox{$\mu<\nu$} and
we check that it is true for at least $\nu$ and $\nu$+1. Let $k$+1 be the level of~$\nu$,
with \hbox{$k\geq 1$} and let $\mu$ be its father which is thus on the level~$k$.

From the statement of the theorem, namely by the rules induced on the types, 
we can deduce the following succession for two nodes~$\kappa$ and~$\kappa$+1 lying on the
same level of the tree.

\vskip 10pt
\ligne{\hfill
$\vcenter{\vtop{\leftskip 0pt\parindent 0pt \hsize=250pt\bf
\ligne{\hfill
\bbzz$,$\wwzu{} $-$ \bbzu$,$\wwuz{} $-$ \wwzz$,$\bbzu{} $-$ \wwuz$,$\bbzz{} $-$ \hfill}
\ligne{\hfill
\wwzu$,$\wwzz{} $-$ \wwuz$,$\wwzz $-$ \wwzu$,$\wwuz\hfill}
}}$
\hfill (4) \hskip 10pt}
\vskip 10pt
Below, Tables~\ref{tb_demsuccb} and~\ref{tb_demsuccw} allow us to check the correctness 
of the statement
of Theorem~\ref{tsuccb}. In both tables, the following notations are used for nodes
connected with $\mu$ or~$\nu$. The node $\mu_1$ is $\mu$$-$1. The node $\nu_1$ is 
$\nu$$-$1. The node $\lambda$ is $s_\ell(\nu)$ and $\lambda_1$ is the node
$\lambda$$-$1. According to~(3), \hbox{$\lambda_1=s_r(\nu_1)$}. In both table,
we indicate the black nodes by writing their number in blue.

In both tables, we apply the induction hypothesis for all nodes~$\kappa$ with
\hbox{$\kappa<\nu$}. In particular, that applies to $\nu_1$ too. Accordingly,
$\lambda_1$, which is always a white node is known. If we know the code of~$\nu_1$,
we know that of~$\lambda_1$, of~$\lambda$
and then those of $\lambda$+1,$\lambda$+2 and $\lambda$+3. 

\def\bbb{{\bf b}}
\def\aa{{\bf a}}
\def\zzz{{\bf 0}}
\def\uu{{\bf 1}}
Indeed, if we know the code of~$\kappa$ we can easily compute the node of \hbox{$\kappa$$-$1}
and of~\hbox{$\kappa$+1}. In $[\kappa]$, say that the letter~\bbb{} is {\bf before} the 
letter~\aa{} if \bbb{} is on the left hand side of~\aa. Say that a \zzz{} is
{\bf safe} if the letter before it is a \zzz{} too. Note that according to~(2), the leftmost
letter of a code is \uu. We assume that there is a safe~\zzz{} before the leading \uu{} of 
the code.

\begin{lemm}\label{linc}
Let $[\kappa]$ be the code of a node of a Fibonacci tree. Let
\hbox{$[\kappa]=[\xi]${\bf 0}$(${\bf 01}$)^n$}. Then,
\hbox{$[\kappa$$+$$1]=[\xi]${\bf 010}$(${\bf 00}$)^{n-1}$}. 
\hbox{$[\kappa]=[\xi]${\bf 00}$(${\bf 10}$)^n$}. Then,
\hbox{$[\kappa$$+$$1]=[\xi]${\bf 01}$(${\bf 00}$)^n$}. We can say that a pattern~\uu\uu{}
generates a carry which propagates until a safe~\zzz{} which it replaces by a~\uu.
\end{lemm}

\begin{lemm}\label{ldec}
Let $[\kappa]$ be the code of a node of a Fibonacci tree. If $[\kappa]$ ends with
a \uu, \hbox{$[\kappa$$-$$1]$} can be written by replacing the rightmost~\uu{} by a~\zzz.
If \hbox{$[\kappa]=[\xi](${\bf 00}$)^n$}. Then,
\hbox{$[\kappa$$-$$1]=[\xi]${\bf 01}$(${\bf 01}$)^{n-1}$}. If
\hbox{$[\kappa]=[\xi]${\bf 0}$(${\bf 00}$)^n$}. Then,
\hbox{$[\kappa$$+$$1]=[\xi]${\bf 0}$(${\bf 10}$)^n$}. We can say that a 
pattern~\zzz\zzz{} generates \uu\zzz{} or \zzz\uu{} depending on the parity
of the repetitions of the pattern~\zzz\zzz{} before the rightmost~\uu. 
If that number is even it generates
\zzz\uu{} for each pattern~\zzz\zzz; otherwise, it generates the pattern~\uu\zzz.
\end{lemm}

\def\lignecalcb #1 #2 #3 #4 #5 #6 #7 #8 
{\scriptsize
\hbox to 20pt{\hfill#1\hfill}
\hbox to 20pt{\hfill#2\hfill}
\hbox to 27pt{\hfill#3\hfill}
\hbox to 27pt{\hfill#4\hfill}
\hbox to 40pt{\hfill#5\hfill}
\hbox to 40pt{\hfill#6\hfill}
\hbox to 40pt{\hfill#7\hfill}
\hbox to 40pt{\hfill#8\hfill}
}

\def\lignecalcw #1 #2 #3 #4 #5 #6 #7 #8 #9 
{\scriptsize
\hbox to 20pt{\hfill#1\hfill}
\hbox to 20pt{\hfill#2\hfill}
\hbox to 27pt{\hfill#3\hfill}
\hbox to 27pt{\hfill#4\hfill}
\hbox to 40pt{\hfill#5\hfill}
\hbox to 40pt{\hfill#6\hfill}
\hbox to 40pt{\hfill#7\hfill}
\hbox to 40pt{\hfill#8\hfill}
\hbox to 40pt{\hfill#9\hfill}
}

\vtop{
\ligne{\hfill
\vtop{\leftskip 0pt\parindent 0pt\hsize=305pt
\begin{tab}\label{tb_demsuccb}
\leurre
Table of the computation of the codes of the following nodes when $\nu$ is a black node:
$\mu$ is the father of~$\nu$, \hbox{$\mu_1=\mu$$-$$1$}, \hbox{$\nu_1=\nu$$-$$1$}, 
$\lambda$, the leftmost son of~$\nu$, \hbox{$\lambda_1=\lambda$$-$$1$}, as well
as the nodes $\lambda$$+$$1$ and $\lambda$$+$$2$. We mark the black nodes in blue,
except when \hbox{$\lambda$$+$$2$} is the successor of~$\nu$.
\end{tab}
}
\hfill}
\vskip 0pt
\ligne{\hfill
\vtop{\leftskip 0pt\parindent 0pt\hsize=311pt
\ligne{\hfill \lignecalcw {$\mu_1$}  {$\mu$} {$\nu_1$} {\Rr{$\nu$}} {$\lambda_1$} 
{\Bb{$\lambda$}} {$\lambda$+1} {\Bb{$\lambda$+2}} {} \hfill}
\vskip 3pt   
\ligne{\hfill $\mu_1$ is a black node, so that $\mu$ is white:
\hfill}
\vskip 3pt
\ligne{\hfill\lignecalcw {\Bb{$\alpha${\bf 00}}} {$\alpha${\bf 01}} 
{$\beta${\bf 1010}} {\Bb{$\alpha$}{\bf \Oo{0000}}} {$\beta${\bf 100100}} 
{\Bb{$\beta${\bf 100101}}} {$\beta${\bf 101010}} {$\alpha${\bf \Oo{0000}\Rr{00}}} 
{\Bb{01},10} \hfill}
\ligne{\hfill\lignecalcw {\Bb{$\beta${\bf 01}}} {$\alpha${\bf 10}} 
{$\beta${\bf 0010}} {\Bb{$\beta$}{\bf \Oo{0100}}} {$\beta${\bf 001000}} 
{\Bb{$\beta${\bf 001001}}} {$\beta${\bf 001010}} {$\beta${\bf \Oo{0100}\Rr{00}}} 
{\Bb{01},10} \hfill}
\vskip 3pt   
\ligne{\hfill $\mu$ is a black node, so that $\mu_1$ is white:
\hfill}
\ligne{\hfill\lignecalcw {$\beta${\bf 10}} {\Bb{$\alpha${\bf 00}}} 
{$\beta${\bf 1000}} {\Bb{$\beta$}{\bf \Oo{1001}}} {$\beta${\bf 100000}} 
{\Bb{$\beta${\bf 100001}}} {$\beta${\bf 100010}} {$\beta${\bf \Oo{1001}\Rr{00}}} 
{\Bb{01},10} \hfill}
\ligne{\hfill\lignecalcw {$\beta${\bf 00}} {\Bb{$\beta${\bf 01}}} 
{$\beta${\bf 0000}} {\Bb{$\beta$}{\bf \Oo{0001}}} {$\beta${\bf 000000}} 
{\Bb{$\beta${\bf 000001}}} {$\beta${\bf 000010}} {$\beta${\bf \Oo{0001}\Rr{00}}} 
{\Bb{01},10} \hfill}
\vskip 3pt   
\ligne{\hfill $\mu_1$ and $\mu$ are both white nodes:
\hfill}
\ligne{\hfill \lignecalcw {$\alpha${\bf 01}} {$\alpha${\bf 10}} 
{$\alpha${\bf 0010}} {\Bb{$\alpha$}{\bf \Oo{0100}}} {$\alpha${\bf 001000}} 
{\Bb{$\alpha${\bf 001001}}} {$\alpha${\bf 001010}} {$\alpha${\bf \Oo{0100}\Rr{00}}} 
{\Bb{01},10} \hfill}
\ligne{\hfill \lignecalcw {$\alpha${\bf 01}} {$\alpha${\bf 00}} 
{$\alpha${\bf 0010}} {\Bb{$\alpha$}{\bf \Oo{0100}}} {$\alpha${\bf 001000}} 
{\Bb{$\alpha${\bf 001001}}} {$\alpha${\bf 001010}} {$\alpha${\bf \Oo{0100}\Rr{00}}} 
{\Bb{01},10} \hfill}
}
\hfill}
}
\vskip 10pt
Our proof consists in carefully looking at every possible case. 
We first assume that $\nu$ is a black node. Its father $\mu$ may be white or black.
Assume that $\mu$$-$1 is black. Necessarily, $\mu$ is white. Now, from the hypothesis,
the possible types for $\mu_1$ and $\mu$, taking into account that 
\hbox{$\mu=\mu_1$+1}, are : \hbox{\bbzz$,$\wwzu} or \hbox{\bbzu$,$\wwuz}. 
As \hbox{$\mu<\nu$}, the induction hypothesis applies to~$\mu$ and to~$\mu_1$,
From~$(4)$, we have that $\mu_1$$-$1 has the type~\wwuz. Accordingly, its rightmost
son ends in \uu\zzz\zzz\zzz, so that as $mu_1$ is a black node, it is easy to check
that $\nu_1$ ends in \uu\zzz\uu\zzz. 
If $\mu$ is black, so that $\mu_1$ is white, the possible succession of the types
are \hbox{\wwzz$,$\bbzu} and \hbox{\wwuz$,$\bbzz}: from the assumption, the rightmost son
of a node is either {\wwuz} or~{\wwzz}. If $\mu_1$ and~$\mu$ are both white,
then the only possibilities for the succession of types are \hbox{\wwzu$,$\wwzz}
and~\hbox{\wwzu$,$\wwuz}. Note that when $\mu_1$ is black or has the type {\wwzu},
its successor is $\nu$, otherwise it is~$\nu_1$. Similarly, if $\nu_1$, which is always
white has the type~{\wwzu}, then its successor is~$\lambda$, otherwise it 
is~$\lambda_1$. From these features and the help of Lemmas~\ref{linc} and~\ref{ldec}
as well as $(4)$,
we obtain the computations of Table~\ref{tb_demsuccb}. Note that in the application
of~\ref{ldec}, we have to take into account on the assumption hypothesis and on the
change from $\beta$01 to $\alpha$00, for instance, in order to compute the
code of~$\kappa$$-$1 when the code of~$\kappa$ ends with several contiguous~\zzz.

Let us apply the similar arguments for the case when $\nu$ is a white node. This time,
the computation is also based on the position of~$\nu$ among the other sons of its 
father~$\mu$. Consider the case when $\mu$ is black. Necessarily, $\mu_1$ is white 
and~$\nu_1$ is black. The possible types for $\mu_1$, $\mu$ and~$\nu_1$ are
\hbox{\wwuz,\bbzz,\bbzu} and \hbox{\wwzz,\bbzu,\bbzu}. In those cases,
the successor of~$\mu_1$ is \hbox{$\nu_2=\nu_1$$-$1}. The computation from~$\nu_2$ 
to~$\nu_1$ is straightforward from Lemma~\ref{linc}, so that we do not mention the code
of~$\nu_2$ in Table~\ref{tb_demsuccw}. Also, as $\nu_1$ has the type~{\bbzu}, its
successor is~$\lambda$.

\vtop{
\ligne{\hfill
\vtop{\leftskip 0pt\parindent 0pt\hsize=305pt
\begin{tab}\label{tb_demsuccw}
\leurre
Table of the computation of the codes of the son of $\nu$ when it is
a white node whose father is $\mu$. The same conventions as in 
Table~{\rm\ref{tb_demsuccb}} are used here too. In the upper part of the table, 
$\mu$ is a black node, so that $\mu_1$ is white.
In the lower part of the table, $\mu$ is a white node and we assume that $\nu_1$ too 
is a white node. In the upper part of the table, $\nu_1$ is black. That node may be also
white in the lower part of the table.
\end{tab}
}
\hfill}
\vskip 0pt
\ligne{\hfill
\vtop{\leftskip 0pt\parindent 0pt\hsize=311pt
\ligne{\hfill $\nu_1$ is black, $\mu_1$ is white and is black:\hfill}
\vskip 3pt
\ligne{\hfill \lignecalcw {$\mu_1$}  {\Bb{$\mu$}} {\Bb{$\nu_1$}} {\Rr{$\nu$}} 
{\Bb{$\lambda$}} {$\lambda$+1} {\Rr{$\lambda$+2}} {\Bb{$\lambda$+3}} {} \hfill}
\vskip 3pt
\ligne{\hfill\lignecalcw {$\beta${\bf 10}} {\Bb{$\alpha${\bf 00}}} 
{\Bb{$\beta${\bf 1001}}} {$\beta${\bf \Oo{1010}}} 
{\Bb{$\beta${\bf 100100}}} {$\beta${\bf 100101}} {$\beta${\bf \Oo{1010}\Rr{00}}} 
{$\beta${\bf 101001}} {\bf \Bb{00},01,00} \hfill} 
\ligne{\hfill\lignecalcw {$\alpha${\bf 00}} {\Bb{$\alpha${\bf 01}}} 
{\Bb{$\alpha${\bf 0000}}} {$\alpha${\bf \Oo{0001}}} 
{\Bb{$\beta${\bf 000000}}} {$\beta${\bf 000001}} {$\beta${\bf 000010}} 
{$\beta${\bf \Oo{0001}\Rr{00}}} {\bf \Bb{00},01,10} \hfill} 
\vskip 3pt
\ligne{\hfill $\nu_1$ and $\mu_1$ are black, $\mu$ is white:\hfill}
\vskip 3pt
\ligne{\hfill \lignecalcw {\Bb{$\mu_1$}}  {$\mu$} {\Bb{$\nu_1$}} {\Rr{$\nu$}} 
{\Bb{$\lambda$}} {$\lambda$+1} {\Rr{$\lambda$+2}} {\Bb{$\lambda$+3}} {} \hfill}
\vskip 3pt
\ligne{\hfill\lignecalcw {\Bb{$\alpha${\bf 01}}} {$\alpha${\bf 10}} 
{\Bb{$\alpha${\bf 0100}}} {$\alpha${\bf \Oo{0101}}} 
{\Bb{$\alpha${\bf 010000}}} {$\alpha${\bf 010001}} {$\alpha${\bf 010010}} 
{$\alpha${\bf \Oo{0101}\Rr{00}}} {\bf \Bb{00},01,10} \hfill} 
\ligne{\hfill\lignecalcw {\Bb{$\alpha${\bf 00}}} {$\alpha${\bf 01}} 
{\Bb{$\alpha${\bf 0000}}} {$\alpha${\bf \Oo{0001}}} 
{\Bb{$\alpha${\bf 000000}}} {$\alpha${\bf 000001}} {$\alpha${\bf 000010}} 
{$\alpha${\bf \Oo{0001}\Rr{00}}} {\bf \Bb{00},01,10} \hfill} 
\vskip 3pt
\ligne{\hfill $\nu_1$ is black, $\mu_1$ and $\mu$ are both white:\hfill}
\vskip 3pt
\ligne{\hfill \lignecalcw {$\mu_1$}  {$\mu$} {\Bb{$\nu_1$}} {\Rr{$\nu$}} 
{\Bb{$\lambda$}} {$\lambda$+1} {\Rr{$\lambda$+2}} {\Bb{$\lambda$+3}} {} \hfill}
\vskip 3pt
\ligne{\hfill\lignecalcw {$\alpha${\bf 01}} {$\alpha${\bf 10}} 
{\Bb{$\alpha${\bf 0100}}} {$\alpha${\bf \Oo{0101}}} 
{\Bb{$\alpha${\bf 010000}}} {$\alpha${\bf 010001}} {$\alpha${\bf 010010}} 
{$\alpha${\bf \Oo{0101}\Rr{00}}} {\bf \Bb{00},01,10} \hfill} 
\ligne{\hfill\lignecalcw {$\beta${\bf 01}} {$\alpha${\bf 00}} 
{\Bb{$\beta${\bf 0100}}} {$\beta${\bf \Oo{0101}}} 
{\Bb{$\beta${\bf 010000}}} {$\beta${\bf 010001}} {$\beta${\bf 010010}} 
{$\beta${\bf \Oo{0101}\Rr{00}}} {\bf \Bb{00},01,10} \hfill} 
\ligne{\hfill\lignecalcw {$\beta${\bf 10}} {$\alpha${\bf 00}} 
{\Bb{$\beta${\bf 1001}}} 
{$\beta${\bf \Oo{1010}}} {\Bb{$\beta${\bf 100100}}} 
{$\beta${\bf 100101}} {$\beta${\bf \Oo{1010}\Rr{00}}} {$\beta${\bf 101001}} 
{\bf \Bb{00},01,00} \hfill}
\vskip 3pt 
\ligne{\hfill $\nu_1$ is a white node, so that $\mu$ is white too:\hfill}
\vskip 3pt 
\ligne{\hfill \lignecalcw {$\mu_1$} {$\mu$} {$\nu_1$} {\Rr{$\nu$}} 
{\Bb{$\lambda$}} {$\lambda$+1} {\Rr{$\lambda$+2}} {\Bb{$\lambda$+3}} {} \hfill}
\vskip 3pt 
\ligne{\hfill\lignecalcw {\Bb{$\alpha${\bf 01}}} {$\alpha${\bf 10}} 
{$\alpha${\bf 0101}} 
{$\alpha${\bf \Oo{1000}}} {\Bb{$\alpha${\bf 010100}}} 
{$\alpha${\bf 010101}} {$\alpha${\bf \Oo{1000}\Rr{00}}} {$\alpha${\bf 100001}} 
{\bf \Bb{00},01,00} \hfill}
\ligne{\hfill\lignecalcw {\Bb{$\alpha${\bf 00}}} {$\alpha${\bf 01}} 
{$\alpha${\bf 0001}} 
{$\alpha${\bf \Oo{0010}}} {\Bb{$\beta${\bf 000100}}} 
{$\beta${\bf 000101}} {$\alpha${\bf \Oo{0010}\Rr{00}}} {$\alpha${\bf 001001}} 
{\bf \Bb{00},01,00} \hfill}
\ligne{\hfill\lignecalcw {$\beta${\bf 01}} {$\alpha${\bf 00}} 
{$\beta${\bf 0101}} 
{$\alpha${\bf \Oo{0000}}} {\Bb{$\beta${\bf 010100}}} 
{$\beta${\bf 010101}} {$\alpha${\bf \Oo{0000}\Rr{00}}} {$\alpha${\bf 000001}} 
{\bf \Bb{00},01,00} \hfill}
\ligne{\hfill\lignecalcw {$\alpha${\bf 01}} {$\alpha${\bf 10}} 
{$\alpha${\bf 0101}} 
{$\alpha${\bf \Oo{1000}}} {\Bb{$\alpha${\bf 010100}}} 
{$\alpha${\bf 010101}} {$\alpha${\bf \Oo{1000}\Rr{00}}} {$\alpha${\bf 000001}} 
{\bf \Bb{00},01,00} \hfill}
\ligne{\hfill\lignecalcw {$\beta${\bf 10}} {$\alpha${\bf 00}} 
{$\beta${\bf 1010}} 
{$\alpha${\bf \Oo{0000}}} 
{\Bb{$\beta${\bf 101001}}} 
{$\beta${\bf 101010}} {$\alpha${\bf \Oo{0000}\Rr{00}}} {$\alpha${\bf 000001}} 
{\bf \Bb{01},10,00} \hfill}
}
\hfill}
}
\vskip 10pt
Consider still the case when $\nu_1$ is black. We have the case when $\mu$ is white and 
both $\mu_1$ is black. The possible types are \hbox{\bbzu,\wwuz,\bbzz} and 
\hbox{\bbzz,\wwzu,\bbzz} for the nodes $\mu_1$, $\mu$ and $\nu_1$ in that order. 
Next, we have the case when both
$\mu_1$ and $\mu$ are white nodes. For $\mu_1$ and~$\mu$
the succession of types is \hbox{\wwzu,\wwuz} or \hbox{\wwzu,\wwzz}
or also \hbox{\wwuz,\wwzz}. In the first two cases, the type of~$\nu_1$ is \bbzz, while in
the exceptional case \hbox{\wwuz,\wwzz}, it is {\bbzu}. 

Now, consider the case when $\nu_1$ is white. Necessarily, $\mu$ is also a white node.
When $\mu_1$ is black, we have the possibilities \hbox{\bbzu,\wwuz} and 
\hbox{\bbzz,\wwzu}. When $\mu_1$ is white, we have three possibilities: 
\hbox{\wwzu,\wwzz} or \hbox{\wwzu,\wwuz} or also
\hbox{\wwzu,\wwzz}. This latter case happens when $\mu$ and $\nu$ are the last nodes 
of two consecutive levels.
We have already met this case when $\mu$ is the last node of a level and $\nu$
is the penultimate node on the next level. In that case $\nu_1$ is black: see 
Table~\ref{tb_demsuccw}. The last column of the tables shows the types of the sons
of $\nu$ with the type written in blue for the black nodes, which allows us to drop the
letter of the status. We can see that the hypothesis is checked for the sons of the
node~$\nu$ and their types. This completes the proof of the theorem. \hfill $\Box$

\subsection{The black Fibonacci tree in the pentagrid and in the heptagrid}
\label{bpentahepta}
  
    It is time to indicate which place a black Fibonacci tree takes in the pentagrid and
in the heptagrid.

    As illustrated by Figure~\ref{fbandes}, the sectors defined by 
Figure~\ref{fspanpentahepta} in Sub section~\ref{tilings} can be split
with the help of regions of the tiling generated by the white Fibonacci tree and by the
black one.

\vskip 10pt
\vtop{
\ligne{\hfill
\includegraphics[scale=0.59]{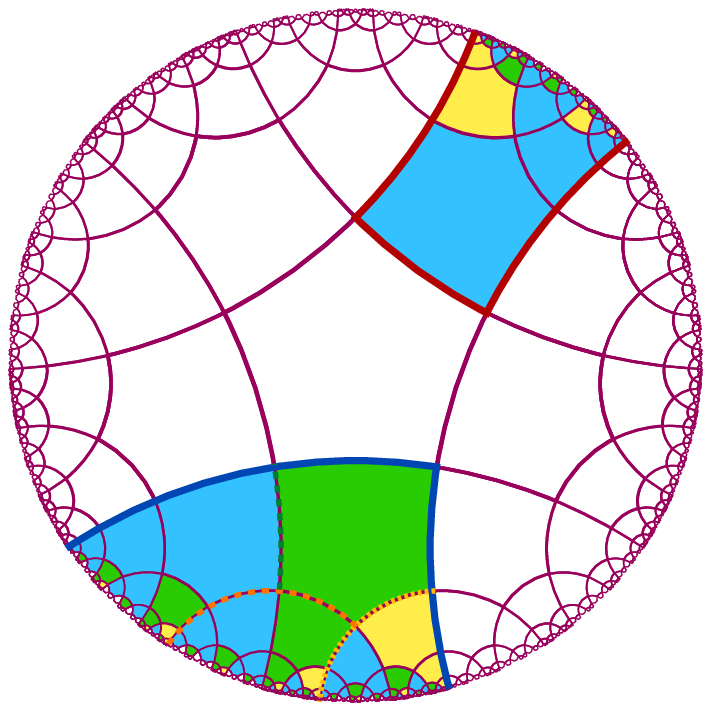}
\includegraphics[scale=0.5]{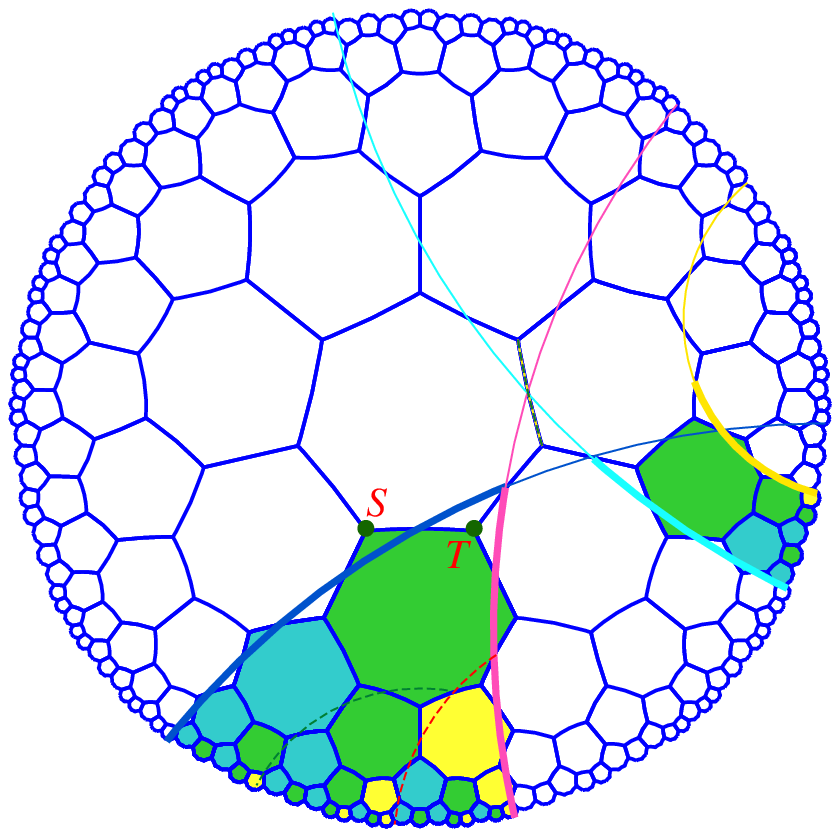}
\hfill}
\vspace{-10pt}
\ligne{\hfill
\vtop{\leftskip 0pt\parindent 0pt\hsize=300pt
\begin{fig}\label{fbandes}
\leurre
The decomposition of a sector spanned by the white Fibonacci tree into a tile, then
two copies of the same sector and a strip spanned by the black Fibonacci tree.
To left: the decomposition in the pentagrid; to right, the decomposition in the heptagrid.
In both cases, the lines which delimit a sector spanned by each kind of tree.
\end{fig}
}
\hfill}
}
\vskip 5pt
In the figure, the sector is split into a tile, we call it the {\bf leading tile}, 
and a complement which can be split into
two copies of the sector and a region spanned by the black Fibonacci tree which we
call a {\bf strip}.

In both tilings, the strip appears as a region delimited by two lines~$\ell_1$
and~$\ell_2$ which are non-secant. It means that they never meet and that they also are 
not parallel, a property which is specific of the hyperbolic plane. There is a third line 
which supports the side of the tile~$\tau$ which is associated with the root of the 
black Fibonacci tree. That line is the common perpendicular to~$\ell_1$ and~$\ell_2$.
The tile~$\tau$ is called the {\bf leading tile} of the strip.
It is worth noticing that the way we used to split the sector can be recursively repeated
in each sector generated by the process of splitting. We can note that the strip itself
can be exactly split into a tile, a sector and a strip. This process is closely related
with the generating rules of the Fibonacci trees. At this point, it can be noticed 
that there are several ways to split a sector and a strip again into strips ans sectors.
This can be associated with other rules for generating a tree which we again call 
a Fibonacci tree. There are still two kinds of nodes, white and black ones. But the
rules are different by the order in which the black son occurs among the sons of a node.
There are two choices for black nodes and three choices for white ones. Accordingly,
there are six possible definitions of Fibonacci tree. We can also decide to
choose which rule is applied each time a node is met. In~\cite{mmJUCStools}
those possibilities are investigated. We refer the interested reader to that paper.

\vskip 10pt
\vtop{
\ligne{\hfill
\includegraphics[scale=0.59]{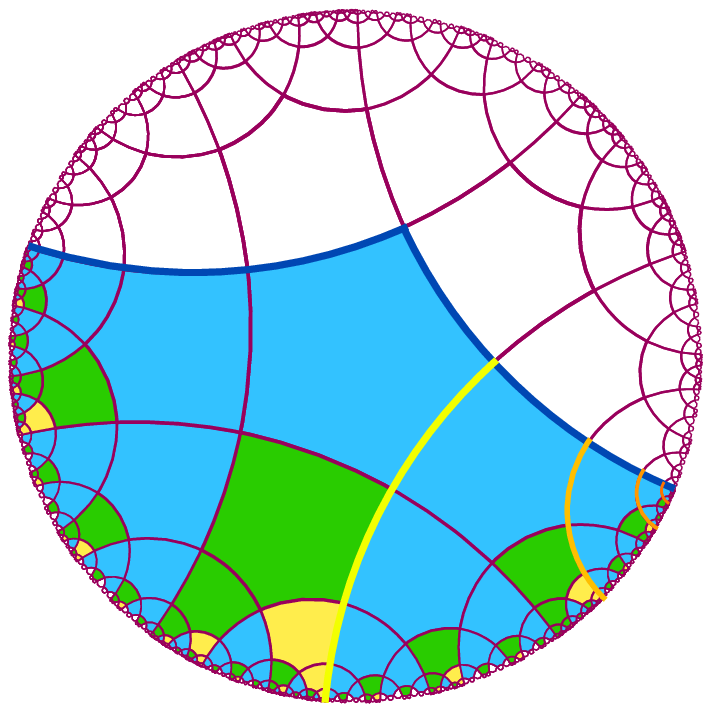}
\includegraphics[scale=0.5]{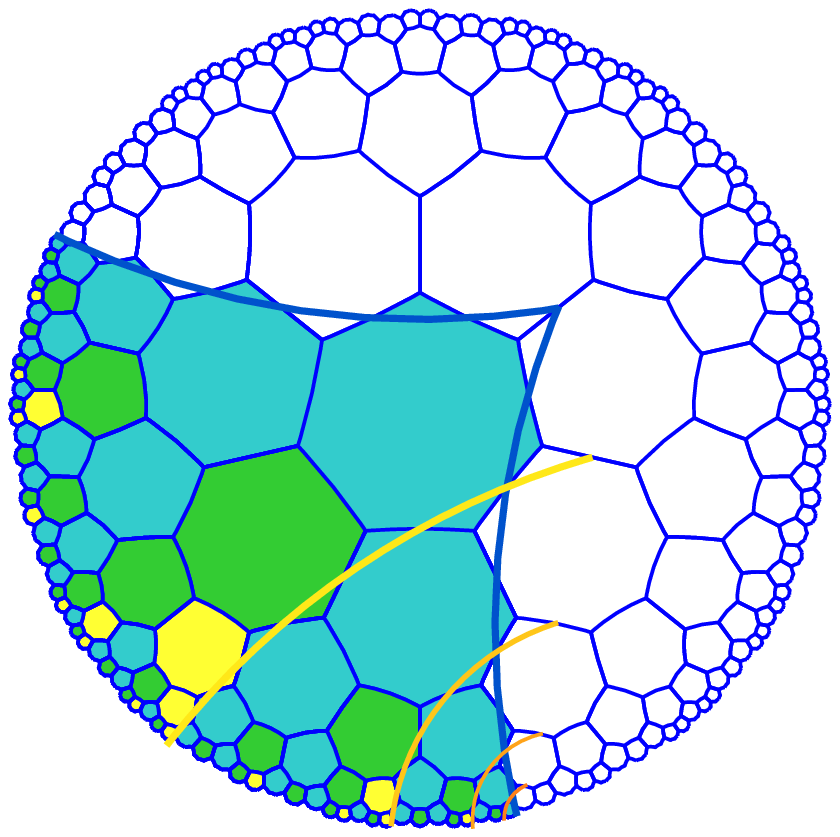}
\hfill}
\vspace{-10pt}
\ligne{\hfill
\vtop{\leftskip 0pt\parindent 0pt\hsize=300pt
\begin{fig}\label{frubans}
\leurre
The decomposition of a sector spanned by the white Fibonacci tree into a sequence of
pairwise adjacent strips spanned by the black Fibonacci tree.
To left: the decomposition in the pentagrid; to right, the decomposition in the heptagrid.
In both cases, the lines which delimit the strips spanned by the tree.
\end{fig}
}
\hfill}
}
\vskip 5pt
\vskip 10pt
\vtop{
\ligne{\hfill
\includegraphics[scale=0.59]{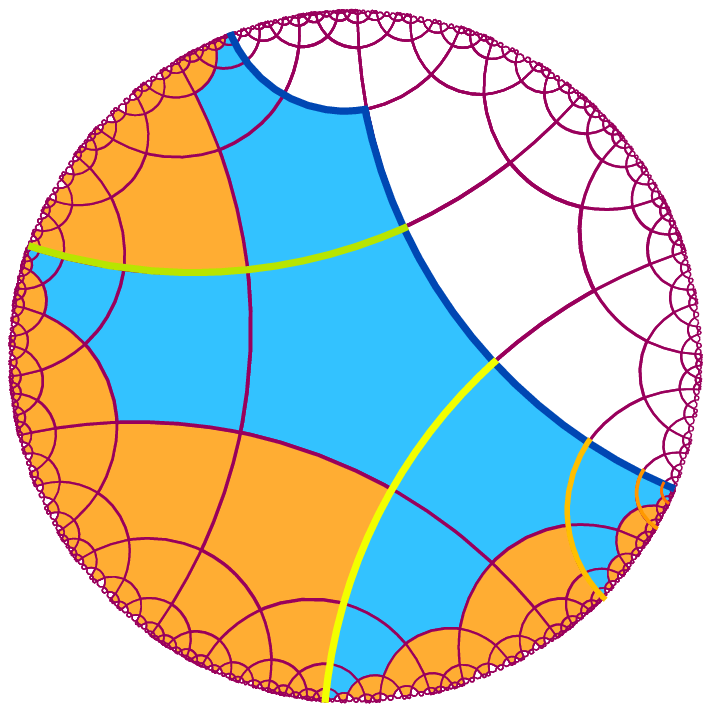}
\includegraphics[scale=0.5]{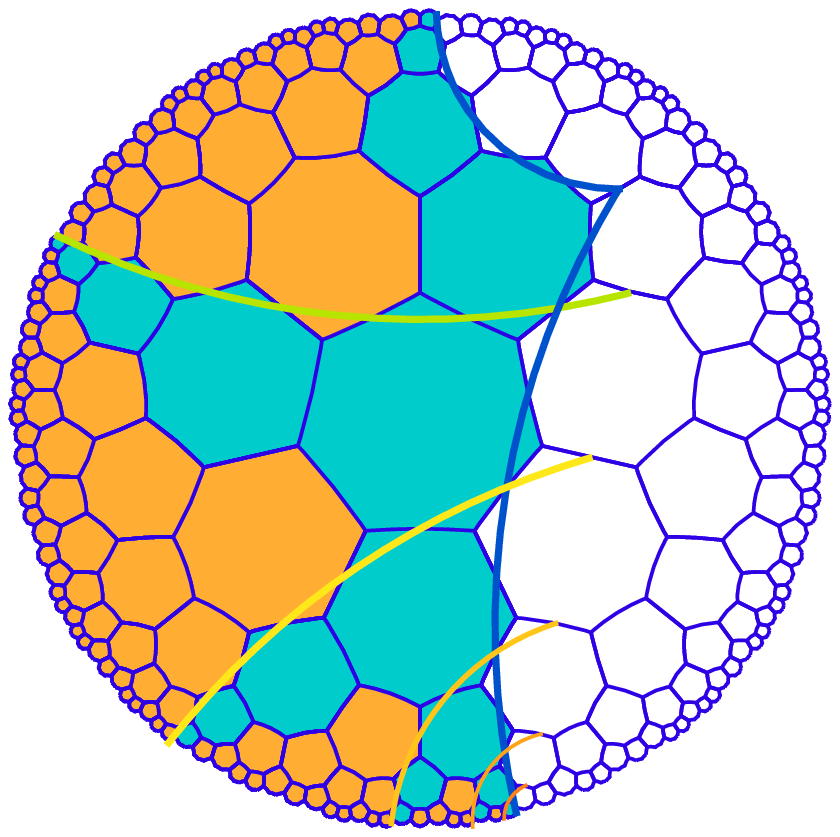}
\hfill}
\vspace{-10pt}
\ligne{\hfill
\vtop{\leftskip 0pt\parindent 0pt\hsize=300pt
\begin{fig}\label{fharpes}
\leurre
Another look on the decomposition of the sector given by Figure~{\rm\ref{frubans}}.
The structure of the Fibonacci is erased in order to highlight the decomposition
into pairwise adjacent strips.
\end{fig}
}
\hfill}
}
\vskip 5pt
   But a sector can be split in another way which is illustrated by 
figure~\ref{frubans}. Consider a sector~${\cal S}_0$. Consider its leading tile~$T$.
That tile is associated with the root of the white Fibonacci tree.
assume that we associate it with the black Fibonacci tree in such a way that
in the association the leftmost son of~$T$ is again the black son of the root in
both trees. What remains in the sector? It remains a node which we can associate
with the root of the white Fibonacci tree. A simple counting argument, taking into
account that the levels are different by one step from the white tree to the black
one in that construction, shows us that in this way we define an exact splitting
of the sector. And so, there is another way to split
the sector: into a strip ${\cal B}_0$ and a sector again, ${\cal S}_1$. Now, what was 
performed 
for~$\cal S$ can be repeated for~${\cal S}_1$ which generates a strip ${\cal B}_1$
and a new sector ${\cal S}_2$. Accordingly,we proved:

\begin{thm}\label{tstrippentahepta}
The sector associated to the white Fibonacci tree can be split into a sequence
of pairwise adjacent strips ${\cal B}_n$, \hbox{$n\in\mathbb N$}, associated to the 
black Fibonacci tree. 
Equivalently, the white Fibonacci tree can be split into the union of a sequence of copies
of the black Fibonacci tree. The leading tiles of the ${\cal B}_n$'s are associated
with the nodes \hbox{$f_{2n+1}$$-$$1$} of the white Fibonacci tree, {\it i.e.} the
nodes which are on the rightmost branch of the white Fibonacci tree.
\end{thm}

\section{When the golden sequence is used}\label{golden}

    Let us go back to the generating rules of the white Fibonacci tree. From the
rules defined by~(1) in Sub section~\ref{fibodef}, we can easily count the number of nodes
which lie at the same level of the tree. Denote by $u_n$, $v_n$ the number of white,
black nodes respectively lying on the level~$n$. Denote the total number of nodes
on that level by $w_n$. Clearly:

\vskip 5pt
\ligne{\hfill
\vtop{\leftskip 0pt\parindent 0pt\hsize=150pt
\ligne{$u_{n+1} = 2u_n+v_n$\hfill}
\ligne{$v_{n+1} = u_n$\hfill}
}
\hfill}   
\vskip 5pt

\noindent
from which we easily get:
\vskip 5pt
\ligne{\hfill $w_{n+2} = 3w_{n+1}-w_n$.
\hfill (5) \hskip 10pt}
\vskip 5pt
 
Note that (5) can be found directly: white nodes generate three nodes and black ones
two nodes only, but the number of black nodes of the considered level is the number
of nodes of the previous level as for each node there is a single black son.
Now, equation~(5) defines a polynomial \hbox{$P(X) = X^2-3X+1$} whose roots are
the numbers 
\hbox{$\displaystyle{{3+\sqrt5}\over2}$} and
\hbox{$\displaystyle{{3-\sqrt5}\over2}$}. Now, 
\hbox{$\displaystyle{{3+\sqrt5}\over2}=\displaystyle{({{1+\sqrt5}\over2}})^2$}
which explains the link with the Fibonacci sequence and why the number of nodes
in the white Fibonacci tree are connected with the Fibonacci numbers whose index is odd. 

We can define codes with the sequence $w_n$. Indeed, it is not difficult to prove
that any positive number~$n$ is a sum of distinct terms of the sequence
\hbox{$\{w_n\}_{n\in\mathbb N^+}$}, where $\mathbb N^+$ is the set of positive integers.
We have:
\vskip 5pt
\ligne{\hfill $n=\displaystyle{\sum\limits_{j=1}^k a_jw_j}$ with $a_j\in\{0,1,2\}$,
\hfill (6)\hskip 10pt}
\vskip 5pt
\noindent
where the sequence \hbox{$\{w_n\}_{n\in\mathbb N^+}$} is defined by~$(5)$ and the initial
conditions \hbox{$w_1=1$} with \hbox{$w_2=3$} for the white Fibonacci tree and
\hbox{$w_1=1$} with \hbox{$w_2=2$} for the black Fibonacci tree.
As in the case of the representation of positive numbers Fibonacci, this representation
is not unique. Also, we can associate to the $a_j$'s of formula~(6) a word in the
alphabet \hbox{$\{0,1,2\}$} which we write \hbox{\bf a$_k$...a$_1$} and which we call
the {\bf golden code} of~$n$ and we write again $[n]$ when there is no ambiguity with
the code defined with the Fibonacci numbers, otherwise we write $[n]_g$ for the golden
code. We can made the golden code of~$n$ unique by requiring that the pattern
\hbox{\bf 21$^*$2} is ruled out, where {\bf 1$^*$} is either the empty word or 
a word consisting of {\bf 1}'s only. The code associated with the Fibonacci numbers
will here be called {\bf Fibonacci code}.

We can also associate the golden code to a tile in a fixed sector of the pentagrid
or of the heptagrid by giving to the tile the golden code of the number associated to
the tile.

   In Sub section~\ref{goldenpreferred}, we look at the properties of the golden code
in the white Fibonacci tree. In Sub section~\ref{goldensucc}, we look at the same issue
in the black Fibonacci tree. We shall see an analogous phenomenon with what we
observed in the previous sections.

\subsection{The golden codes in the white Fibonacci trees}\label{goldenpreferred}

   The golden codes in the white Fibonacci tree have properties which are similar to those
of the Fibonacci codes we have depicted in Sub section~\ref{preferred} in the same tree.

   The author and its co-author, Gencho Skordev, proved that the preferred son property
is true with the golden codes. However, due to the different alphabet used for writing
the codes, the definition of the preferred son in this context is a bit different.
The definition comes with the following result:

\begin{thm} {\rm (see \cite{mmgsJUCS,mmbook1})}\label{tgoldenw}
In the white Fibonacci tree fitted with the golden codes, each node has exactly one node
among its sons whose code ends with \zzz. That son is called the {\bf preferred son} of
the node. Moreover, if $[\nu]$ is the golden code of~$\nu$, the golden code of its
preferred son is $[nu]$\zzz. In all nodes, the preferred son is the leftmost white son.
\end{thm}

\vskip 10pt
\vtop{
\ligne{\hfill
\includegraphics[scale=0.35]{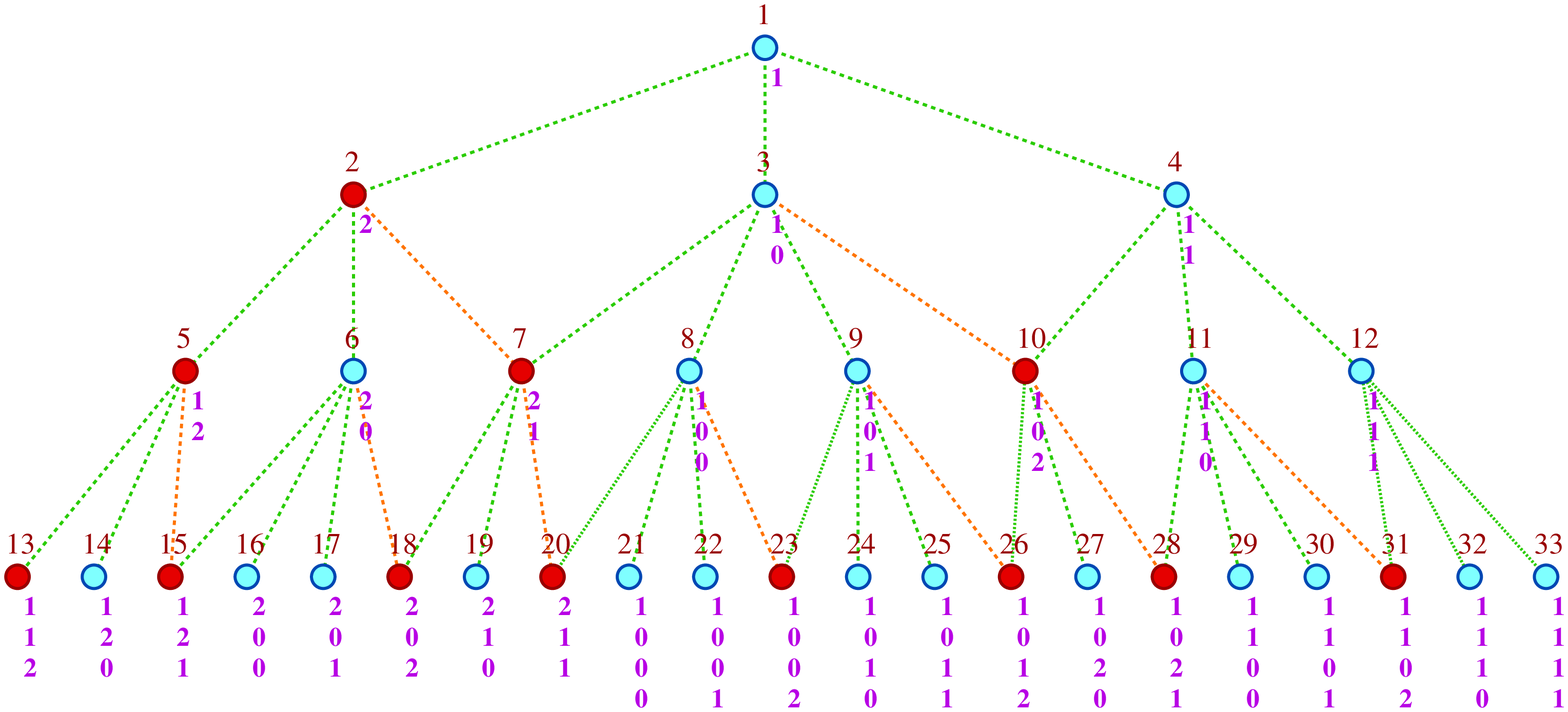}
\hfill}
\vspace{-15pt}
\ligne{\hfill
\vtop{\leftskip 0pt\parindent 0pt\hsize=300pt
\begin{fig}\label{fgoldenw}
\leurre
The golden codes in the white Fibonacci tree: we can check the properties stated
by Theorem~{\rm\ref{tgoldenw}}.
\end{fig}
}
\hfill}
}

We refer the reader to~\cite{mmgsJUCS,mmbook1} for the proof of the result which is stated
for the general case in the quoted references.

\subsection{The golden codes in the black Fibonacci trees}\label{goldensucc}

   In this context too, when we look at the golden codes in the black Fibonacci tree,
we can see that the preferred son property is no more true, as it can easily be
checked on Figure~\ref{fgoldenb}.

   However, there is a kind of regularity, more regular than in the case of the
Fibonacci codes, which are indicated in Theorem~\ref{tgoldenb}. As in
Sub section~\ref{bsucc}, we call {\bf successor} of the node~$\nu$, the node whose
golden code is $[\nu]_g$\zzz, the node being again denoted by $succ(\nu)$.

\def\bgz{\hbox{\bf b0}}
\def\bgu{\hbox{\bf b1}}
\def\wgz{\hbox{\bf w0}}
\def\wgu{\hbox{\bf w1}}
\def\wgd{\hbox{\bf w2}}

\begin{thm}\label{tgoldenb}
In the black Fibonacci tree, for all nodes~$\nu$, we have that, except for the white nodes 
whose golden code ends in~\zzz, 
\hbox{$succ(\nu)=s_r(\nu)$$+$$1$}. For the white nodes whose golden code ends in~\zzz,
we have \hbox{$succ(\nu)=s_r(\nu)$}. Moreover, the last two letters of a golden codes
combined with the statuses of the nodes give rise to five combinations:
\bgz, \bgu, \wgz, \wgu{} and \wgd. Each type give rises to a specific
rule determining the types of the sons:
\vskip 5pt
\ligne{\hfill
\vtop{\leftskip 0pt\parindent 0pt\hsize=100pt
\ligne{\bgz{} $\rightarrow$ \bgz$,$ \wgu\hfill}
\ligne{\bgu{} $\rightarrow$ \bgu$,$ \wgd\hfill}
\ligne{\wgz{} $\rightarrow$ \bgz$,$ \wgu$,$ \wgz\hfill}
\ligne{\wgu{} $\rightarrow$ \bgz$,$ \wgu$,$ \wgd\hfill}
\ligne{\wgd{} $\rightarrow$ \bgz$,$ \wgu$,$ \wgd\hfill}
}
\hfill}
\end{thm}
The proof is very similar in its principle to the proof of Theorem~\ref{tsuccb}. This is
why, here, it will boil down to Tables~\ref{tb_demgoldsuccw} and~\ref{tb_demgoldsuccb}.
We just append the following remark: the successive types of two consecutive nodes
are the following ones, which is a consequence of the rules and which are used also 
among the induction hypothesis:
\vskip 10pt
\ligne{\hfill
$\vcenter{\vtop{\leftskip 0pt\parindent 0pt\hsize=250pt\bf 
\ligne{\hfill \bgz$,$\wgu{} $-$ \bgu$,$\wgd{} $-$ \wgu$,$\bgz{} $-$ \hfill}
\ligne{\hfill \wgd$,$\bgz{} $-$ \wgu$,$\wgz{} $-$ \wgu$,$\wgd \hfill}
}}$
\hfill (7) \hskip 10pt}
\vskip 10pt

Indeed, we shall check by induction that the case \hbox{\wgz$,$\bgu} never occurs
for nodes on the same level as a second white son in \zzz{} occurs only for the last
node of a level.

 Before turning to the tables, we can check the properties stated in Theorem~\ref{tgoldenb}
in Figure~\ref{fgoldenb}. In the tables, the notations for the numbers of the nodes
are the same as in Tables~\ref{tb_demsuccb} and~\ref{tb_demsuccw}. We have again,
that \hbox{$\lambda=s_\ell(\nu)=s_r(\nu_1)$}, and that \hbox{$\nu_1=s_r(\mu_1)$}.

\def\dd{{\bf 2}}
    Although the general splitting of the proof is the same as in the case of the
Fibonacci codes, there are in this case specificities connected with the special forbidden
pattern \hbox{\dd$^\ast$\dd} and with the fact that a node of type~{\wgz} has its rightmost
son as its successor. We also have to check that such a node occurs as the last node of
the level only. Also, the succession of types for two nodes is ruled by~$(7)$, which
is different from~$(4)$ in the codes based on Fibonacci numbers. The situation of the
white node ending with~\zzz{} appears in Table~\ref{tb_demgoldsuccw} as the third line
of the case when $\nu_1$ is also white. We can see that a code \hbox{\bf$\beta$11}
is followed by the code \hbox{\bf $\alpha$00}, which means that $\beta$ ends with a
\dd{} followed by \uu's only and so that $\alpha$ is the successor of~$\beta$.
This explains why the code of $\lambda$+1, \hbox{\bf$\beta$111} is followed
by the code \hbox{\bf $\alpha$000} which is that of $\lambda$+2 as expected from
the statement of Theorem~\ref{tgoldenb} as in this case, the type of $\nu$ is {\wgz}.
From this and from the table, we can see that a rightmost son of type {\wgz} occurs
at this line only, so that the assumption that the last node on a level is of that type
is checked.

\vtop{
\begin{tab}\label{tb_demgoldsuccb}
\leurre
Table of the computation of the golden codes of the following nodes when $\nu$ is a 
black node. We mark the black nodes in blue,
except when \hbox{$\lambda$$+$$2$} is the successor of~$\nu$.
\end{tab}

\ligne{\hfill
\vtop{\leftskip 0pt\parindent 0pt\hsize=311pt
\ligne{\hfill \lignecalcw {$\mu_1$}  {$\mu$} {$\nu_1$} {\Rr{$\nu$}} {$\lambda_1$} 
{\Bb{$\lambda$}} {$\lambda$+1} {\Bb{$\lambda$+2}} {} \hfill}
\vskip 3pt   
\ligne{\hfill $\mu_1$ is a black node, so that $\mu$ is white:
\hfill}
\vskip 3pt
\ligne{\hfill\lignecalcw {\Bb{$\alpha${\bf 0}}} {$\alpha${\bf 1}} 
{$\beta${\bf 12}} {\Bb{$\alpha$}{\bf \Oo{00}}} {$\beta${\bf 112}} 
{\Bb{$\beta${\bf 120}}} {$\beta${\bf 121}} {$\alpha${\bf \Oo{00}\Rr{0}}} 
{\Bb{0},1} \hfill}
\ligne{\hfill\lignecalcw {\Bb{$\alpha${\bf 0}}} {$\alpha${\bf 1}} 
{$\beta${\bf 11}} {\Bb{$\alpha$}{\bf \Oo{00}}} {$\beta${\bf 102}} 
{\Bb{$\beta${\bf 110}}} {$\beta${\bf 111}} {$\alpha${\bf \Oo{00}\Rr{0}}} 
{\Bb{0},1} \hfill}
\ligne{\hfill\lignecalcw {\Bb{$\alpha${\bf 1}}} {$\alpha${\bf 2}} 
{$\alpha${\bf 02}} {\Bb{$\alpha$}{\bf \Oo{10}}} {$\alpha${\bf 012}} 
{\Bb{$\alpha${\bf 020}}} {$\alpha${\bf 021}} {$\alpha${\bf \Oo{10}\Rr{0}}} 
{\Bb{0},1} \hfill}
\vskip 3pt   
\ligne{\hfill $\mu_1$ is a white node and $\mu$ is black:
\hfill}
\vskip 3pt   
\ligne{\hfill\lignecalcw {$\beta${\bf 1}} {\Bb{$\alpha${\bf 0}}} 
{$\beta${\bf 02}} {\Bb{$\beta$}{\bf \Oo{10}}} {$\beta${\bf 012}} 
{\Bb{$\beta${\bf 020}}} {$\beta${\bf 021}} {$\beta${\bf \Oo{10}\Rr{0}}} 
{\Bb{0},1} \hfill}
\ligne{\hfill\lignecalcw {$\beta${\bf 2}} {\Bb{$\alpha${\bf 0}}} 
{$\beta${\bf 12}} {\Bb{$\beta$}{\bf \Oo{20}}} {$\beta${\bf 112}} 
{\Bb{$\beta${\bf 120}}} {$\beta${\bf 121}} {$\beta${\bf \Oo{20}\Rr{0}}} 
{\Bb{0},1} \hfill}
\vskip 3pt   
\ligne{\hfill $\mu_1$ and $\mu$ are white:
\hfill}
\vskip 3pt   
\ligne{\hfill\lignecalcw {$\beta${\bf 1}} {$\alpha${\bf 0}} 
{$\beta${\bf 02}} {\Bb{$\beta$}{\bf \Oo{10}}} {$\beta${\bf 012}} 
{\Bb{$\beta${\bf 020}}} {$\beta${\bf 021}} {$\beta${\bf \Oo{10}\Rr{0}}} 
{\Bb{0},1} \hfill}
\ligne{\hfill\lignecalcw {$\alpha${\bf 1}} {$\alpha${\bf 2}} 
{$\alpha${\bf 02}} {\Bb{$\alpha$}{\bf \Oo{10}}} {$\alpha${\bf 012}} 
{\Bb{$\alpha${\bf 020}}} {$\alpha${\bf 021}} {$\alpha${\bf \Oo{10}\Rr{0}}} 
{\Bb{0},1} \hfill}
}
\hfill}
}
\vskip 10pt
\vtop{
\ligne{\hfill
\includegraphics[scale=0.35]{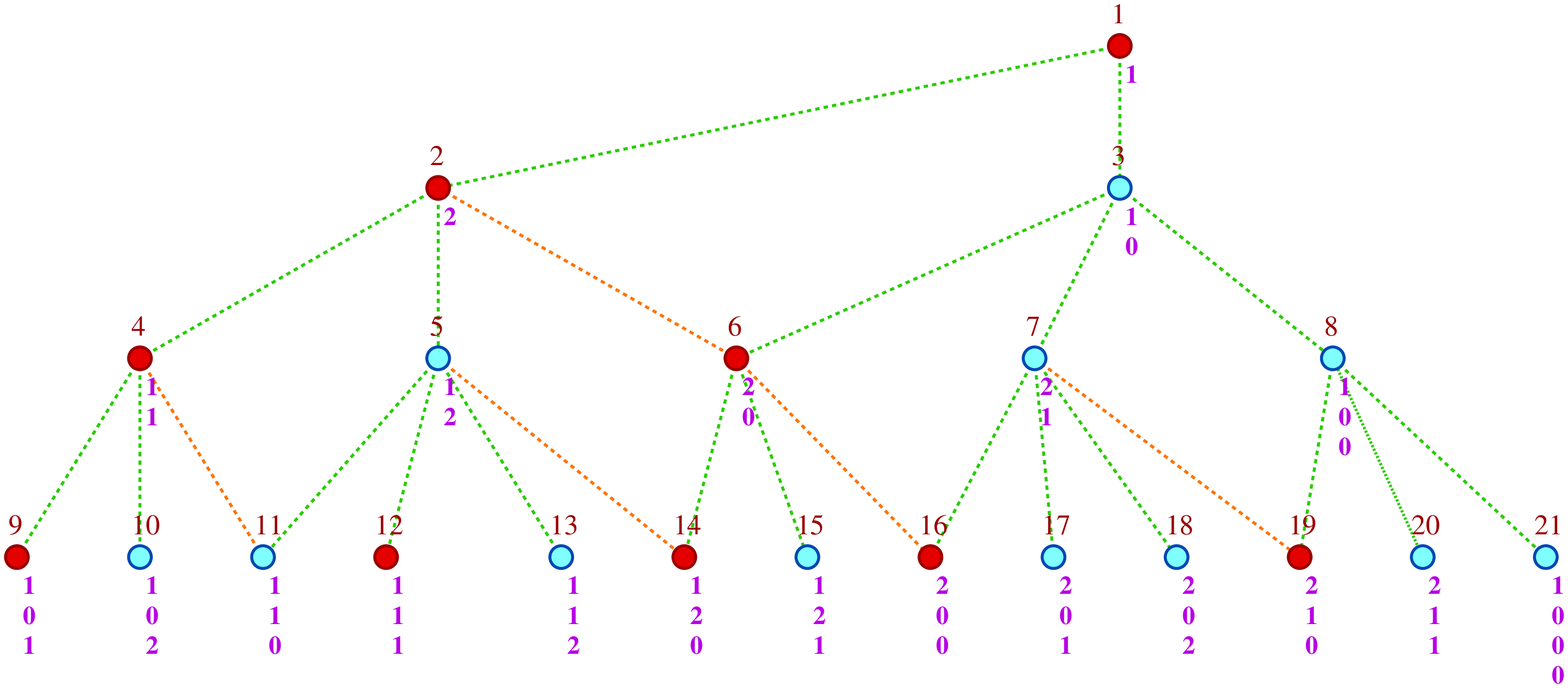}
\hfill}
\vspace{-25pt}
\ligne{\hfill
\vtop{\leftskip 0pt\parindent 0pt\hsize=300pt
\begin{fig}\label{fgoldenb}
\leurre
The golden codes in the black Fibonacci tree: we can check the properties stated
by Theorem~{\rm\ref{tgoldenb}}.
\end{fig}
}
\hfill}
}
\vskip 5pt
Also note that in Table~\ref{tb_demgoldsuccb}, the first two lines of the table
indicates two different situations with the predecessor of a code \hbox{$\alpha$\zzz}.
In the first line, we have the standard situation where the predecessor is 
\hbox{$\beta$\dd}.
In the second line, the predecessor is \hbox{$\beta$\uu}. This means that 
in that case, $\beta$ ends in a pattern \hbox{\dd\uu$^\ast$}, so that the next node
on the level has the code \hbox{$\alpha$\zzz$^\ast$}. Also note that when
the code of the node~$\kappa$ is $\beta$\uu{} and that of~$\kappa$+1 is
$\alpha$\zzz, it means that $\beta$ ends with the pattern \hbox{\bf 21$^\ast$},
so that the code $\beta$\uu$^k$ is followed by $\alpha$\zzz$^\ast$. In the same line,
if the code of~$\kappa$ is $\alpha$\uu{} and that of~$\kappa$+1 is
$\alpha$\dd, it means that $\alpha$ does not end with \hbox{\bf 21$^\ast$}, so that
the code $\alpha$\uu$^{k+1}$ is followed by $\alpha$\uu$^k$\dd.

\vskip 5pt

\vtop{
\begin{tab}\label{tb_demgoldsuccw}
\leurre
Table of the computation of the golden codes of the following nodes when $\nu$ is a 
white node. We mark the black nodes in blue,
except when \hbox{$\lambda$$+$$2$} is the successor of~$\nu$.
\end{tab}
\ligne{\hfill
\vtop{\leftskip 0pt\parindent 0pt\hsize=311pt
\ligne{\hfill $\nu_1$ and $\mu$ are black nodes, so that $\mu_1$ is white: \hfill}
\vskip 3pt
\ligne{\hfill \lignecalcw {$\mu_1$}  {\Bb{$\mu$}} {\Bb{$\nu_1$}} {\Rr{$\nu$}} 
{\Bb{$\lambda$}} {$\lambda$+1} {$\lambda$+2} {\Bb{$\lambda$+3}} {} \hfill}
\vskip 3pt   
\ligne{\hfill\lignecalcw {$\beta${\bf 1}} {\Bb{$\alpha${\bf 0}}} 
{\Bb{$\beta${\bf 10}}} {$\beta${\bf \Oo{11}}} {$\beta${\bf 100}} 
{\Bb{$\beta${\bf 101}}} {$\beta${\bf 102}} {$\beta${\bf \Oo{11}\Rr{0}}} 
{\Bb{0},1} \hfill}
\ligne{\hfill\lignecalcw {$\beta${\bf 2}} {\Bb{$\alpha${\bf 0}}} 
{\Bb{$\beta${\bf 20}}} {$\beta${\bf \Oo{21}}} {$\beta${\bf 200}} 
{\Bb{$\beta${\bf 201}}} {$\beta${\bf 202}} {$\beta${\bf \Oo{21}\Rr{0}}} 
{\Bb{0},1} \hfill}
\vskip 3pt   
\ligne{\hfill $\nu_1$ is a black node, and $\mu$ is white: \hfill}
\vskip 3pt
\ligne{\hfill \lignecalcw {$\mu_1$}  {$\mu$} {\Bb{$\nu_1$}} {\Rr{$\nu$}} 
{\Bb{$\lambda$}} {$\lambda$+1} {$\lambda$+2} {\Bb{$\lambda$+3}} {} \hfill}
\vskip 3pt   
\ligne{\hfill\lignecalcw {\Bb{$\alpha${\bf 0}}} {$\alpha${\bf 1}} 
{\Bb{$\alpha${\bf 00}}} {$\alpha${\bf \Oo{01}}} {$\alpha${\bf 000}} 
{\Bb{$\alpha${\bf 001}}} {$\alpha${\bf 002}} {$\alpha${\bf \Oo{01}\Rr{0}}} 
{\Bb{0},1,2} \hfill}
\ligne{\hfill\lignecalcw {\Bb{$\alpha${\bf 1}}} {$\alpha${\bf 2}} 
{\Bb{$\alpha${\bf 10}}} {$\alpha${\bf \Oo{11}}} {$\alpha${\bf 100}} 
{\Bb{$\alpha${\bf 101}}} {$\alpha${\bf 102}} {$\alpha${\bf \Oo{11}\Rr{0}}} 
{\Bb{0},1,2} \hfill}
\ligne{\hfill\lignecalcw {$\beta${\bf 1}} {$\alpha${\bf 0}} 
{\Bb{$\beta${\bf 10}}} {$\beta${\bf \Oo{11}}} {$\beta${\bf 100}} 
{\Bb{$\beta${\bf 101}}} {$\beta${\bf 102}} {$\beta${\bf \Oo{11}\Rr{0}}} 
{\Bb{0},1,2} \hfill}
\ligne{\hfill\lignecalcw {$\alpha${\bf 1}} {$\alpha${\bf 2}} 
{\Bb{$\alpha${\bf 10}}} {$\alpha${\bf \Oo{11}}} {$\alpha${\bf 100}} 
{\Bb{$\alpha${\bf 101}}} {$\alpha${\bf 102}} {$\alpha${\bf \Oo{11}\Rr{0}}} 
{\Bb{0},1,2} \hfill}
\vskip 3pt   
\ligne{\hfill $\nu_1$ is a white node, and so $\mu$ is white too: \hfill}
\vskip 3pt
\ligne{\hfill \lignecalcw {$\mu_1$}  {$\mu$} {$\nu_1$} {\Rr{$\nu$}} 
{\Bb{$\lambda$}} {$\lambda$+1} {$\lambda$+2} {\Bb{$\lambda$+3}} {} \hfill}
\vskip 3pt   
\ligne{\hfill\lignecalcw {\Bb{$\alpha${\bf 0}}} {$\alpha${\bf 1}} 
{$\alpha${\bf 01}} {$\alpha${\bf \Oo{02}}} {$\alpha${\bf 010}} 
{\Bb{$\alpha${\bf 011}}} {$\alpha${\bf 012}} {$\alpha${\bf \Oo{02}\Rr{0}}} 
{\Bb{0},1,2} \hfill}
\ligne{\hfill\lignecalcw {\Bb{$\alpha${\bf 1}}} {$\alpha${\bf 2}} 
{$\alpha${\bf 11}} {$\alpha${\bf \Oo{12}}} {$\alpha${\bf 110}} 
{\Bb{$\alpha${\bf 111}}} {$\alpha${\bf 112}} {$\alpha${\bf \Oo{12}\Rr{0}}} 
{\Bb{0},1,2} \hfill}
\ligne{\hfill\lignecalcw {$\beta${\bf 1}} {$\alpha${\bf 0}} 
{$\beta${\bf 11}} {$\alpha${\bf \Oo{00}}} {$\beta${\bf 110}} 
{\Bb{$\beta${\bf 111}}} {$\alpha${\bf \Oo{00}\Rr{0}}} {$\alpha${\bf 001}} 
{\Bb{0},1,0} \hfill}
\ligne{\hfill\lignecalcw {$\alpha${\bf 1}} {$\alpha${\bf 2}} 
{$\alpha${\bf 11}} {$\alpha${\bf \Oo{12}}} {$\alpha${\bf 110}} 
{\Bb{$\alpha${\bf 111}}} {$\alpha${\bf 112}} {$\alpha${\bf \Oo{12}\Rr{0}}} 
{\Bb{0},1,2} \hfill}
}
\hfill
}
}
\vskip 10pt
With those last remarks and Table~\ref{tb_demgoldsuccw}, we completed
the proof of Theorem~\ref{tgoldenb}.\hfill $\Box$

\section{Conclusion}\label{conclude}

\section*{Conclusion}

    We can conclude the paper with several remarks.

    The first one is the interest of the golden representation, which is not
used as intensively as it should be by the author itself, although he found out
the property stated by Theorem~\ref{tgoldenw} a long time ago. The property
stated in Theorem~\ref{tgoldenb} is rather unexpected. It is a new one and it explains
the weak use of that encoding. 

    The second remark, connected with both the Fibonacci and the golden representations 
is that the white Fibonacci tree is the best tree for navigation purpose in the pentagrid
and in the heptagrid, those tessellation that live in the hyperbolic plane. 

    A third remark is that it was proved in \cite{mmgsJUCS,mmbook1}, that the properties
found out in the pentagrid and in the heptagrid can be generalized to the
tessellations \hbox{$\{p,4\}$} and \hbox{$\{p$+$2,3\}$}, still in the hyperbolic plane,
where \hbox{$p\geq5$}. As in the case of the pentagrid and of the heptagrid which 
corresponds to the case \hbox{$p=5$}, for each $p$, a specific tree generates the
tiling in both \hbox{$\{p,4\}$} and \hbox{$\{p$+$2,3\}$}. Interestingly, the rules
associated with such a tree extend in some sens the rules used to construct the
Fibonacci trees. And so, in that generalized context, the white and the black trees
also exist. Now, as the preferred son property is also true in the white tree, this 
time for an extension of the golden encoding, we can wonder whether an extension
of Theorem~\ref{tgoldenb} is true for the black tree.

\end{document}